\documentclass[journal,10pt]{IEEEtran}
\usepackage[utf8]{inputenc}
\usepackage[T1]{fontenc}
\usepackage{amsmath,amssymb,amsfonts,times}
\usepackage{graphicx}
\usepackage{textcomp}
\usepackage{pgfplots}
\usepackage{scalefnt}

\newcommand{\norm}[1]{\left\lVert#1\right\rVert}
\newcommand{\abs}[1]{\left\vert#1\right\vert}
\newcommand{\diag}{\mathop{\rm diag}}

\begin{document}

\title{Study of Switched Max-Link Relay Selection for Cooperative Multiple-Antenna Systems\\
}

\author{F. L. Duarte
            and R. C. de Lamare,~\IEEEmembership{Senior Member,~IEEE}
\thanks{Copyright (c) 2015 IEEE. Personal use of this material is permitted. However, permission to use this material for any other purposes must be obtained from the IEEE by sending a request to pubs-permissions@ieee.org.}
\thanks{F. L. Duarte is with the Centre for Telecommunications Studies (CETUC), Pontifical Catholic University of Rio de Janeiro, Brazil, and the Military Institute of Engineering, IME, Rio de Janeiro, RJ, Brazil. e-mail: flaviold@cetuc.puc-rio.br}
\thanks{R. C. de Lamare is with the Centre for Telecommunications Studies (CETUC), Pontifical Catholic University of Rio de Janeiro, Brazil, and the Department of Eletronic Engineering, University of York, United Kingdon. e-mail: delamare@cetuc.puc-rio.br}}

\maketitle
\begin{abstract}
In this work, we present a switched relaying framework for multiple-input multiple-output (MIMO) relay systems where a source node may transmit directly to a destination node or aided by relays. We also investigate relay selection techniques for the proposed switched relaying framework, whose relays are equipped with buffers. In particular, we develop a novel relay selection protocol based on switching and the selection of the best link, denoted as Switched Max-Link. We then propose the Maximum Minimum Distance (MMD) relay selection criterion for MIMO systems, which is based on the optimal Maximum Likelihood (ML) principle and can provide significant performance gains over other criteria, along with algorithms that are incorporated into the proposed Switched Max-Link protocol. An analysis of the proposed Switched Max-Link protocol and the MMD relay selection criterion in terms of computational cost, pairwise error probability, sum-rate and average delay is carried out. Simulations show that Switched Max-Link using the MMD criterion outperforms previous works in terms of sum-rate, pairwise error probability, average delay and bit error rate.
\end{abstract}

\begin{IEEEkeywords}
Cooperative communications, Relay-selection, Max-Link, Maximum Likelihood criterion, MIMO
\end{IEEEkeywords}

\IEEEpeerreviewmaketitle

\section{INTRODUCTION}
\IEEEPARstart{I}{n} wireless networks, signal fading caused by multipath propagation
is a channel propagation phenomenon that can be mitigated through the use of cooperative
diversity\cite{f1,f2,f3}.  In
cooperative communications with multiple relays, where a number of
relays help a source to transmit data packets to a destination,
by receiving, decoding and forwarding these packets,
relay selection schemes are key because of their high performance \cite{f5, f4, f35}.
As cooperative communication can improve the throughput and extend
the coverage of wireless communications systems, the task of relay
selection  serves as a building block to realize it. In this context, relay
schemes have been included in recent/future wireless
standards such as Long Term Evolution (LTE) Advanced \cite{f6, f13} and
5G standards \cite{f12}.

\subsection{Prior and Related Work}

 In conventional relaying, using half duplex (HD) and
decode-and-forward protocols, transmission is often organized in a
prefixed schedule with two successive time slots. In the first time
slot, the relay receives and decodes the data transmitted from the
source, and in the second time slot the relay forwards the decoded
data to the destination.  Single relay selection schemes use the
same relay for reception and transmission, and cannot simultaneously exploit the best available source-relay ($SR$) and
relay-destination ($RD$) channels. The most common schemes are
bottleneck based and maximum harmonic mean based best relay
selection (BRS) \cite{f5}.

The performance of relaying schemes can be improved if
the link with the highest power is used in each time slot.
This can be achieved via a buffer-aided relaying protocol, where
the relay can accumulate packets in its buffer prior to transmission.
The use of buffers provides an improved
performance and extra degrees of freedom for system design \cite{f6, f14}. However,
it suffers from additional delay that must
be well managed for delay-sensitive applications. Buffer-aided
relaying protocols require not only the acquisition of channel state
information (CSI), but control of the buffer status. Applications of buffer-aided relaying are: vehicular,
cellular, and sensor networks \cite{f6}.

In Max-Max Relay Selection (MMRS) \cite{f5}, in the first time slot,
the relay selected for reception can store the received packets in
its buffer and forward them at a later time when selected for
transmission. In the second time slot, the relay selected for
transmission can transmit the first packet in the queue of its
buffer, which was received from the source earlier. MMRS  assumes
infinite buffer sizes. However, considering finite buffer sizes, the buffer of a relay becomes empty if the
channel conditions are such that it is selected repeatedly for
transmission (and not for reception) or full if it is selected
repeatedly for reception (and not for transmission). To overcome
this limitation, in  \cite{f5} a hybrid relay selection (HRS) scheme, which is a combination of  BRS and MMRS,
was proposed.

Although MMRS and HRS  improve the throughput and/or SNR gain as
compared to BRS, their diversity gain is limited to the number
of relays $N$. This can be improved by combining adaptive link
selection with MMRS, which results in the
Max-Link \cite{f9} protocol. The main idea of Max-Link is to select in each
time slot the strongest link among all the available $SR$ and $RD$ links
(i.e., among $2N$ links) for transmission \cite{f7}. For independent and identically distributed (i.i.d.)  links and no
delay constraints, Max-Link achieves a diversity gain of $2N$, which
is twice the diversity gain of BRS and MMRS.

 Max-Link has been extended in \cite{f11} to account for direct
source-destination ($S D$) connectivity, which provides resiliency
in low transmit SNR conditions \cite{f7}. In
\cite{f19,f20,f21,f22,f23,f24,f25}, buffer-aided relay selection
protocols were shown to improve the Max-Link performance by reducing
the average packet delay, ensuring a good diversity gain, and/or
achieving full diversity gain with a smaller buffer size as compared
to Max-Link. In \cite{f19}, the outage performance and the average
packet delay of a  relay system that exploits buffer-aided max-link
relay selection are analyzed. In \cite{f20}, a study of the average
packet delay of a buffer-aided scheme that selects a relay node
based on both the channel quality and the buffer state of the relay
nodes was performed. In \cite{f21}, the relay associated  with the
largest weight is selected among the qualified source-relay and
relay-destination links, where each link is assigned with a weight
related to the buffer status. In \cite {f22}, motivated by the
Max-Link  and the Max-Max protocols, a hybrid buffer-aided
cooperative protocol that attains the benefits of reliability and
reduced packet delay is reported. In \cite{f23}, a delay and
diversity-aware buffer-aided relay selection policy that reduces the
average delay and obtains a good diversity gain is proposed. In
\cite{f24}, a relay selection scheme that seeks to maintain the
states of the buffers by balancing the arrival and departure rates
at each relay's buffer has been reported. In \cite{f25}, the best
relay node is selected as the link with the highest channel gain
among the links within a priority class. In summary, the previous
schemes (MMRS, HRS and Max-Link) only use buffer-aided relay
selection for cooperative single-antenna systems.

More recently, buffer-aided relay selection protocols for
cooperative multiple-antenna systems have been studied. In
\cite{456}, a virtual full-duplex (FD) buffer-aided relaying to
recover the loss of multiplexing gain caused by HD relaying in a
multiple relay network through joint opportunistic relay selection
(RS) and beamforming (BF), is presented. Moreover, in \cite{654}, a
cooperative network with a buffer-aided multi-antenna source,
multiple HD buffer-aided relays and a single destination is
presented to recover the multiplexing loss of the network.

\subsection{Contributions}

In this work, we develop a switched relaying framework extended  for
MIMO relay systems that considers direct or cooperative
transmissions with Maximum Likelihood (ML) detection and a Switched
Max-Link  protocol for cooperative MIMO systems, with non reciprocal
channels, which selects the best links among $N$ relay nodes and
whose preliminary results were reported in \cite{f27} and then
further detailed in \cite{smax_link}. We then consider the novel MMD
relay selection criterion \cite{f27}, which is based on the optimal
ML principle and the Pairwise Error Probability (PEP)
\cite{f27,f78,f87}, and the existing Quadratic Norm (QN) criterion
and devise relay selection algorithms for Switched Max-Link. An
analysis of the proposed scheme in terms of PEP, sum-rate, average
delay and computational cost is also carried out. Simulations
illustrate the excellent performance of the proposed framework, the
proposed Switched Max-Link protocol and the MMD-based relay
selection algorithm as compared to previously reported approaches.
The main contributions of this work can be summarized as:
\begin{enumerate}
\item A  switched relaying framework extended for MIMO relay systems that considers direct or cooperative transmissions with ML detection;
 \item The Switched  Max-Link protocol for cooperative MIMO relay systems;
\item The MMD criterion for MIMO relay systems, along with a relay selection algorithm;
\item An analysis of the proposed Switched Max-Link scheme with the MMD relay selection criterion in terms of PEP, sum-rate, average delay and computational cost.
\end{enumerate}

Table \ref{table1} shows the description of the main symbols adopted in this work.

\begin{table}[!htb]
\centering
 \caption{Description of the symbols}
 \label{table1}
\begin{tabular}{l|ll}
\hline
Symbols& Description\\
\hline
$D$ & Destination node\\
\hline
$\mathcal{D}$& MMD metric\\
\hline
$\mathcal{D}_{\min}$& Minimum distance \\
\hline
$\mathcal{D}'_{\min}$& Minimum value of the PEP argument\\
\hline
$d_c$ & Distances between the constellation symbols\\
\hline
$E[d_n]^{MMD}$& Average delay of the MMD-Max-Link protocol\\
\hline
$E[d_n]^{SML}$& Average delay of the Switched Max-Link protocol\\
\hline
$E[L_n]$ & Average queue length\\
\hline
$E_S$ & Energy transmitted from $S$\\
\hline
$E_{R_j}$ & Energy transmitted from $R_j$\\
\hline
$E[T_n]$ & Average throughput of a relay\\
\hline
$\mathbf{H}_{S,D}$ & Matrix of $SD$ links\\
\hline
$\mathbf{H}_{S,R_k}$ & Matrix of $SR_k$ links\\
\hline
$\mathbf{H}^u_{S,R_k}$ & Submatrix of $SR_k$ links\\
\hline
$\mathbf{H}_{R_j,D}$ & Matrix of $R_jD$ links\\
\hline
$\mathbf{H}^u_{R_j,D}$ & Submatrix of $R_jD$ links\\
\hline
$J$ & Size of the buffer (in packets)\\
\hline
$L$ & Queue length\\
\hline
$M_S$ & Number of antennas at $S$ and $D$\\
\hline
$M_R$ & Number of antennas at the relays\\
\hline
$N$ & Number or relays\\
\hline
$N_s$ & Number of constellation symbols\\
\hline
$N_0$ & Power spectral density of the AWGN\\
\hline
$\mathbf{n}_{D}$ & AWGN at $D$\\
\hline
$\mathbf{n}_{R_k}$ & AWGN at $R_k$\\
\hline
$P_{ML}^{\mathcal{S}}$ & Probability of operating in the Max-Link mode\\
\hline
$\mathcal{Q}$& QN metric\\
\hline
$\mathbf{Q}_{S,D}$ & Covariance matrix of the transmitted symbols (for $SD$)\\
\hline
$\mathbf{Q}_{S,R_k}$ & Covariance matrix of the transmitted symbols (for $SR_k$)\\
\hline
$\mathbf{Q}_{R_j,D}$ & Covariance matrix of the transmitted symbols (for $R_jD$)\\
\hline
$R_k$ & Relay selected for reception\\
\hline
$R_j$ & Relay selected for transmission\\
\hline
$\mathcal{R}$& Sum-Rate\\
\hline
$S$ & Source node\\
\hline
$\mathcal{S}$ & Switch of the Switched Max-Link protocol\\
\hline
$U$ & Number of sets of $M_S$ antennas at the relays\\
\hline
$\mathbf{x}$ & Vector of transmitted symbols\\
\hline
$\mathbf{\hat{x}}$ & Estimate of the vector of transmitted symbols\\
\hline
$\mathcal{X}$ & Number of calculations of the MMD metric\\
\hline
$\mathbf{y}_{S,D}$ & Received vector of symbols (for $SD$ links)\\
\hline
$\mathbf{y}_{S,R_k}$ & Received vector of symbols (for $SR_k$ links)\\
\hline
$\mathbf{y}_{R_j,D}$ & Received vector of symbols (for $R_jD$ links)\\
\hline
$\rho$ & Average data rate\\
\hline

\end{tabular}
\end{table}

This paper is structured as follows. Section II describes the system
model and the main assumptions made. Section III details the proposed Switched
Max-Link protocol with the MMD relay selection criterion whereas Section IV analyzes it.
Section V illustrates and discusses the numerical results whereas Section VI gives the
concluding remarks.

\section{System Description}

\begin{figure}[!h]
\centering
\includegraphics[scale=0.6]{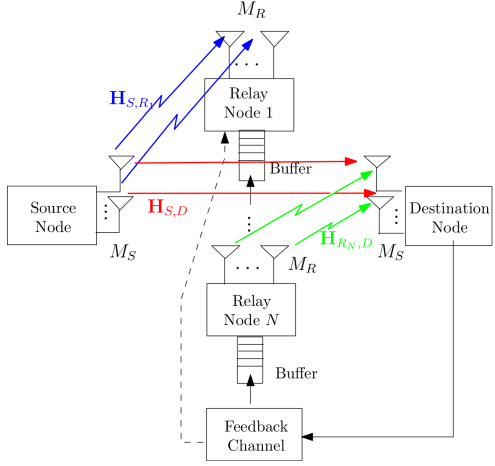}
\caption{System Model}
\label{fig:model}
\end{figure}

We consider a multiple-antenna relay network with one source node, $S$, one
destination node, $D$, and $N$ half-duplex decode-and-forward (DF)
relays, $R_1$,...,$R_N$.  The $S$ and $D$ nodes have $M_S$ antennas for transmission and reception, respectively, and each relay $M_R=U M_S$ antennas, where $U\in \{1,2,3\dots\}$.  All the $M_R$ antennas are used for reception ($M_{R_{rx}}=M_R$) and a set of $M_S$ antennas is selected among $M_R$ to be used for transmission ($M_{R_{tx}}=M_S$). Thus, this configuration forms a spatial multiplexing network, in which the channel matrices are square or formed by multiple square submatrices. Each relay is equipped with a buffer, whose size is $J$ packets and the
transmission is organized in time slots  \cite{f5}. This configuration is considered for simplicity. The considered
system is shown in Fig. \ref{fig:model}.

\subsection{Assumptions}

In cooperative transmissions two time slots are needed to transmit
data packets from $S$ to $D$, so the energy
transmitted in direct transmissions (from $S$ to $D$) is twice the energy $E_S$ transmitted in cooperative
transmissions, from $S$ to the relay selected for reception $R_k$  or
from the relay selected for transmission $R_j$ to $D$ ($E_{R_j}$),  $E_{R_j}=E_S=E$. For
this reason, the energy transmitted from each antenna in cooperative transmissions equals $E/M_S$ and the energy transmitted from each antenna in direct transmissions equals $2 E/M_S$. We consider that the channel coefficients are modeled by mutually independent
zero mean complex Gaussian random variables.
Moreover, we assume that the transmission is organized in data
packets and the channels are constant for the duration of one time slot
and vary independently from one time slot to the next. The information about the order of the data packets is contained in
the preamble of each packet, so the original order is restored at
$D$. Other information such as signaling for CSI estimation are also inserted in the preamble of the packet. We consider perfect and imperfect CSI.  A distributed implementation can reduce signaling overheads and reduce the impact of outdated CSI.
Furthermore, we assume that the relays do not communicate with each
other. We also assume that $D$ is the central node, being responsible for deciding whether $S$ or a relay should transmit in a given time slot $i$. The central node has
access to the channel and the buffer state information, so it may run the algorithm in each time slot and select the
relay for transmission or reception through a
feedback channel. This assumption can be ensured by an
appropriate signalling that provides global CSI at $D$ \cite{f9}. Furthermore, we assume that $S$ has no CSI and each relay has only information about its $SR$ channels and  buffer status.

\subsection{System Model}

The proposed system can operate in each time slot in two modes: "Direct Transmission" (DT)  or "Max-Link". Thus, depending on the relay selection metrics (explained in Section III), the system may operate in each time slot with three options:

a) DT mode: $S$ transmits $M_S$ packets directly to $D$;

b) Max-Link-$SR$ mode: $S$ transmits $M_S$  packets to $R_k$;

c)  Max-Link-$RD$ mode: $R_j$ transmits $M_S$ packets to $D$.
\\

If the relay selection algorithm decides to operate in the DT mode, the received signal from the $S$ to $D$ is
organized in an $M_S \times 1$ vector $\mathbf{y}_{S,D}[i]$ given
by
\begin{eqnarray}
    \mathbf{y}_{S,D}[i]= \sqrt{\frac{2 E}{M_S}} \mathbf{H}_{S,D}\mathbf{x}[i]+\mathbf{n}_D[i],
    \label{eq:1}
\end{eqnarray}
\noindent where $\mathbf{x}[i]$ represents
the vector formed by $M_S$ symbols sent
by $S$, $\mathbf{H}_{S,D}$  represents the $M_S \times  M_S$ matrix of $SD$
links and $\mathbf{n}_D$  denotes the zero mean additive white
complex Gaussian noise (AWGN) at $D$.
Assuming synchronization and perfect CSI, at $D$ we employ the ML receiver which yields
    \begin{eqnarray}
    \hat{\mathbf{x}}[i]= \arg \min_{\mathbf{x'}[i]} \left(\norm{\mathbf{y}_{S,D}[i]- \sqrt{\frac{2 E}{M_S}} \mathbf{H}_{S,D}\mathbf{x'}[i]}^2\right),
    \label{eq:5}
    \end{eqnarray}
where $\mathbf{x'}[i]$ represents each possible vector formed by
$M_S$ symbols. Thus, the ML receiver computes the vector of
transmitted symbols which is the optimal solution. As an example, if
we have BPSK (number of constellation symbols $N_s= 2$), unit power
symbols and $M_S = 2$, the estimated vector of transmitted symbols $
\hat{\mathbf{x}}[i]$ may be $[-1~-1]^T$, $[-1~+1]^T$, $[+1~-1]^T$ or
$[+1~+1]^T$. Other suboptimal detection techniques could be
considered in future work
\cite{mmimo,wence,deLamare2003,itic,deLamare2008,cai2009,jiomimo,Li2011,dfcc,deLamare2013,did,rrmser,bfidd,1bitidd,aaidd}.

Otherwise, if the relay selection algorithm decides to operate in
the Max-Link-$SR$ mode, the received signal from $S$ to $R_k$ is
organized in an $UM_S \times 1$ vector $\mathbf{y}_{S,R_k }[i]$
given by
\begin{eqnarray}
    \mathbf{y}_{S,R_k}[i]=\sqrt{\frac{E}{M_S}} \mathbf{H}_{S,R_k}\mathbf{x}[i]+\mathbf{n}_{R_k}[i],
    \label{eq:2}
\end{eqnarray}
\noindent where
$\mathbf{H}_{S,R_k}$ represents the $UM_S \times  M_S$ matrix of $S R_k$ links and
$\mathbf{ n}_{R_k}$ represents the AWGN at $R_k$. Note that $\mathbf{H}_{S,R_k}$ is formed by $U$ square submatrices of dimensions $M_S \times M_S$ as given by
 \begin{eqnarray}
\mathbf{H}_{S,R_k}= [\mathbf{H}^1_{S,R_k}; \mathbf{H}^2_{S,R_k}; \dots ; \mathbf{H}^U_{S,R_k}].
 \end{eqnarray}

Assuming synchronization and perfect CSI, at  $R_k$ we employ the ML receiver \cite{f4}:
    \begin{eqnarray}
    \hat{\mathbf{x}}[i]= \arg \min_{\mathbf{x'}[i]} \left(\norm{\mathbf{y}_{S,R_k}[i]- \sqrt{\frac{E}{M_S}} \mathbf{H}_{S,R_k}\mathbf{x'}[i]}^2\right).
    \label{eq:4}
    \end{eqnarray}

Moreover,  if the relay selection algorithm decides to operate in the Max-Link-$RD$ mode, the signal transmitted from $R_j$ to $D$ is structured in an $M_S \times 1$ vector
$\mathbf{y}_{R_j,D }[i]$ given by
    \begin{eqnarray}
    \mathbf{y}_{R_j,D}[i]=\sqrt{\frac{E}{M_S}}  \mathbf{H}^u_{R_j,D}\hat{\mathbf{x}}[i]+\mathbf{n}_D[i],
    \label{eq:3}
    \end{eqnarray}
\noindent where $\hat{\mathbf{x}}[i]$ is the vector formed by $M_S$
previously decoded symbols in the relay selected for reception and
stored in its buffer and now transmitted by $R_j$ and
$\mathbf{H}^u_{R_j,D}$ is an $M_S \times M_S$ matrix of  $R_jD$
links. Alternatively, a designer can consider precoding techniques
\cite{lclattice,switch_int,switch_mc,gbd,wlbd,mbthp,rmbthp,bbprec,baplnc}
to help mitigate interference rather than open loop transmission.
Note that $\mathbf{H}^u_{R_j,D}$ is selected among $U$ submatrices
of dimension $M_S \times M_S$ contained in $\mathbf{H}_{R_j,D}$ as
given by
 \begin{eqnarray}
\mathbf{H}_{R_j,D}= [\mathbf{H}^1_{R_j,D}; \mathbf{H}^2_{R_j,D}; \dots ; \mathbf{H}^U_{R_j,D}].
 \end{eqnarray}

At $D$, we also resort to the ML receiver which computes
    \begin{eqnarray}
    \hat{\mathbf{x}}[i]= \arg \min_{\mathbf{x'}[i]} \left(\norm{\mathbf{y}_{R_j,D}[i]- \sqrt{ \frac{E}{M_S}} \mathbf{H}^u_{R_j,D}\mathbf{x'}[i]}^2\right).
    \label{eq:6}
    \end{eqnarray}
Considering imperfect CSI, the estimated channel matrix
$\mathbf{\hat{H}}$ is assumed, instead of $\mathbf{H}$ in
(\ref{eq:5}), (\ref{eq:4})  and (\ref{eq:6}): a channel error matrix
$\mathbf{H}_e$ is added to the channel matrix ($\mathbf{H}_{S,R_k}$,
$\mathbf{H}_{R_j,D}$ or $\mathbf{H}_{S,D}$) and we focus on the case
where errors decay as $O (SNR^{-\alpha})$ for some constant
$\alpha\in [0,1]$  \cite{f16}. Thus,  the variance of the
$\mathbf{H}_e$ coefficients is given by $\sigma_e^2=\beta
E^{-\alpha}$ ($\beta \geq 0$), in the case of $\mathbf{H}_{S,R_k}$
or $\mathbf{H}_{R_j,D}$, and $\sigma_e^2=\beta (2 E)^{-\alpha}$,  in
the case of $\mathbf{H}_{S,D}$.  As an example, in the case of
$\mathbf{H}_{S,R_k}$, the estimated channel matrix is given by
$\mathbf{\hat{H}}_{S,R_k} = \mathbf{H}_{S,R_k}  + \mathbf{H}_e$.
Channel and parameter estimation
\cite{smce,TongW,jpais_iet,armo,badstc,baplnc,goldstein,qian,jio,jidf,jiols,jiomimo}
techniques could be considered in future work in order to develop
algorithms for this particular setting.

\section{Principles of Switched Max-Link Relay Selection Based on MMD}

In this section, we detail the proposed Switched Max-Link relay
selection protocol.

 \subsection{Principles of Switched Max-Link Relay Selection}

The system presented in Fig. \ref{fig:model} is equipped with the proposed Switched Max-Link relay selection protocol, that in each time slot may operate in two possible modes ("DT" or
"Max-Link"), with three options:

a) work in DT mode: $S$ sends $M_S$ packets directly to $D$;

b) work in Max-Link-$SR$ mode: $S$ sends $M_S$
packets to $R_k$ and these packets are
stored in its buffer;

c) work in Max-Link-$RD$ mode: $R_j$
forwards $M_S$ packets from its buffer to $D$.

The proposed Switched Max-Link protocol uses the MMD relay selection criterion. As the scheme proposed in \cite{f10}, the proposed MMD relay selection
 criterion is based on the ML principle. However, the metrics calculated by MMD are different from those of the scheme in \cite{f10}, which leads to considerably better performance. MMD is also based on the worst case of the  PEP and chooses the relay associated with the largest minimum Euclidian distance. So, it requires the distance between the $N_s^{M_S}$
possible vectors of transmitted symbols.  The MMD-based relay selection algorithm, in the Max-Link-$SR$ mode, chooses the relay $R_k$ and the associated channel matrix $\mathbf{H}_{S,R_k}^{MMD}$ with the largest minimum distance as given by
\begin{eqnarray}
\mathbf{H}_{S,R_k}^{MMD}=\arg \max_{\mathbf{H}_{S,R_i}} \mathcal{D}_{\min SR_i},
  \label{eq:78}
\end{eqnarray}
where $\mathcal{D}_{\min SR_i}=\min \left(\frac{E}{M_S}\norm{\mathbf{H}_{S,R_i}^u(\mathbf{x}_l -
\mathbf{x}_n)}^2\right)$, $u \in \{1, \dots U\}$, $i \in \{1, \dots N\}$, $\mathbf{x}_l$  and $\mathbf{x}_n$ represent each possible vector formed by $M_S$ symbols and $l$ $\neq$ $n$. The metric $\frac{E}{M_S}\norm{\mathbf{H}_{S,R_i}^u(\mathbf{x}_l -
\mathbf{x}_n)}^2$  is calculated for each of the $C_2^{N_s^{M_S}}$
(combination of $N_s^{M_S}$  in $2$) possibilities, for each submatrix $\mathbf{H}_{S,R_i}^u$, and $\mathcal{D}_{\min SR_i}$ is the smallest of these values, for each $R_i$. Thus, the selected matrix $\mathbf{H}_{S,R_k}^{MMD}$ has the largest  $\mathcal{D}_{\min SR_i}$ value.

Moreover, the MMD-based relay selection algorithm, in the Max-Link-$RD$ mode, chooses the relay $R_j$ and the associated channel matrix $\mathbf{H}_{R_j,D}^{MMD}$ with the largest minimum distance as given by
\begin{eqnarray}
\mathbf{H}_{R_j,D}^{MMD}=\arg \max_{\mathbf{H}_{R_i,D}} \mathcal{D}_{\min R_iD},
   \label{eq:788}
\end{eqnarray}
where $\mathcal{D}_{\min R_iD} = \max{(\mathcal{D}^u_{\min R_iD})}$ and $\mathcal{D}^u_{\min R_iD}=   \min \left(\frac{E}{M_S}\norm{\mathbf{H}^u_{R_i,D}(\mathbf{x}_l -
\mathbf{x}_n)}^2\right)$.  Note that the submatrix $\mathbf{H}^u_{R_j,D}$ associated with  the largest  $\mathcal{D}^u_{\min R_iD}$ value  is selected among $U$ submatrices of dimension $M_S \times M_S$ contained in $\mathbf{H}_{R_j,D}^{MMD}$.
Table \ref{table3} shows the Switched Max-Link pseudo-code and the following
subsections explain how this protocol works.

\begin{table}[!htb]
\centering
 \caption{Switched Max-Link  Pseudo-Code}
 \label{table3}
\begin{tabular}{l}

\hline
\\
1:   ~~~~ Calculate the metrics $\mathcal{D}^u_{SR_i}$, of each submatrix $\mathbf{H}^u_{S,R_i}$ of  $R_i$ \\
       ~~~~~~      $\mathcal{D}^u_{SR_i}=  \norm{\sqrt{ E/M_S} \mathbf{H}^u_{S,R_i}\mathbf{x}_l - \sqrt{ E/M_S} \mathbf{H}^u_{S,R_i}\mathbf{x}_n}^2$ ; \\
~~~~~~~~~ ~~~~~~~~~~~~~~~~~~~~~~~~~~~~~~~~~~~~~~~~~~~~~~~~ $ i=1,...,N$\\
~~~~~~~~~ ~~~~~~~~~~~~~~~~~~~~~~~~~~~~~~~~~~~~~~~~~~~~~~~ $ u =1,...,U$\\
~~~~~~~~~ ~~~~~~~~~~~~~~~~~~~~~~~~~~~~~~~~~~~~~~~~~~~~~~~~~$ l= 1,...,N_s^{M_S} - 1$\\
~~~~~~~~~ ~~~~~~~~~~~~~~~~~~~~~~~~~~~~~~~~~~~~~~~~~~~~~~~ $ n = l+1,...,N_s^{M_S}$ \\

2:  ~~~~  Find the minimum distance - $\mathcal{D}^u_{\min SR_i}$\\
~~~~~~    $ \mathcal{D}^u_{\min SR_i} = \min{(\mathcal{D}^u_{SR_i})};$ \\
\\

3:   ~~~~ Calculate the metrics $\mathcal{D}^u_{R_iD}$,  of each submatrix $\mathbf{H}^u_{R_i,D}$ of  $R_i$\\
   ~~~~~~      $\mathcal{D}^u_{R_iD}=  \norm{\sqrt{ E/M_S} \mathbf{H}^u_{R_i,D}\mathbf{x}_l - \sqrt{ E/M_S} \mathbf{H}^u_{R_i,D}\mathbf{x}_n}^2$ ; \\
\\
4:  ~~~~  Find the minimum distance - $\mathcal{D}^u_{\min R_iD}$\\
  ~~~~~~    $ \mathcal{D}^u_{\min R_iD} = \min{(\mathcal{D}^u_{R_iD})};$ \\\\

5:  ~~~~  Find the largest minimum distance - $\mathcal{D}_{\min R_iD}$\\
  ~~~~~~    $ \mathcal{D}_{\min R_iD} = \max{(\mathcal{D}^u_{\min R_iD})};$ \\
\\
6: ~~~~Compute the expected values and  $\mathcal{D}_{\min SR_i}$ \\
  ~~~~~~ $\mathcal{D}_{\min SR_i}= \frac{\mbox{E} [\mathcal{D}_{\min R_iD}]}{\mbox{E} [\mathcal{D}^u_{\min SR_i}]}\mathcal{D}^u_{\min SR_i};$\\\\

7:  ~~~~  Perform ordering on $\mathcal{D}_{\min SR_i}$ and $\mathcal{D}_{\min R_iD}$\\
\\
8:  ~~~~   Find the maximum minimum distance\\
~~~~~~ $\mathcal{D} _{\max \min SR-RD}= \max{(\mathcal{D}_{\min SR_i},\mathcal{D}_{\min R_iD})}$;
\\
\\
9:   ~~~~ Calculate the metrics $\mathcal{D}_{SD}$\\

       ~~~~~~      $\mathcal{D}_{SD}=  \norm{\sqrt{2 E/M_S} \mathbf{H}_{S,D}\mathbf{x}_l - \sqrt{ 2 E/M_S} \mathbf{H}_{S,D}\mathbf{x}_n}^2;$ \\
\\
10:  ~~~~  Find the minimum distance - $\mathcal{D}_{\min SD}$\\
 ~~~~~~    $ \mathcal{D}_{\min SD} = \min{(\mathcal{D}_{SD})};$ \\

\\

11:  ~~~~  Select the transmission mode\\
 ~~~~~~ $\mathcal{D} _{\max \min}=\mathcal{D} _{\max \min SR-RD};$\\
 ~~~~~~   $G=\frac{\mathcal{D}_{\max\min}}{\mathcal{D}_{\min SD}}$;\\
\\
$\mbox{Mode}=
\begin{cases}
  \mbox{Max-Link-}\textit{SR},& \mbox{if}\left(\mathcal{D} _{\max \min}= \max{(\mathcal{D}_{\min SR_i})}\right)\& \left(G> \mathcal{S}\right), ^\dagger\\
    \mbox{Max-Link-}\textit{RD},&\mbox{if}\left(\mathcal{D} _{\max \min}= \max{(\mathcal{D}_{\min R_i D})}\right)\& \left( G> 1\right),\\
 \mbox{DT},&\mbox{otherwise.}
            \end{cases}
$\\\\
~~~~~~$^\dagger$ Note that $\mathcal{S} \in \{0,1,2,...\}$ is a parameter that works as a \\
~~~~~~~~switch. When $\mathcal{S}=0$ the scheme operates only in the Max-Link-\\
 ~~~~~~~ ($SR$ or $RD$)  mode (MMD-Max-Link protocol). Moreover, when  \\
 ~~~~~~~  $\mathcal{S}>0$  the scheme operates  in the Max-Link-($SR$ or $RD$)  \\
~~~~~~~  or DT mode  (Switched-Max-Link protocol).
\\\\
\hline
\end{tabular}
\end{table}

\subsection{Calculation of relay selection metric}

In the first step we calculate the metrics $\mathcal{D}^u_{SR_i}$
related to the $SR$ channels of each submatrix $\mathbf{H}^u_{S,R_i}$ of  each relay $R_i$, in Max-Link mode:
\begin{eqnarray}
  \mathcal{D}^u_{SR_i}=  \norm{\sqrt{ \frac{E}{M_S}} \mathbf{H}^u_{S,R_i}\mathbf{x}_l - \sqrt{ \frac{E}{M_S}} \mathbf{H}^u_{S,R_i}\mathbf{x}_n}^2,
  \label{eq:7}
\end{eqnarray}
where $u \in \{1, \dots U\}$, $i \in \{1, \dots N\}$,  $\mathbf{x}_l$  and $\mathbf{x}_n$
represent each possible vector formed by $M_S$ symbols and $l$ $\neq$ $n$. This metric is calculated for each of the $C_2^{N_s^{M_S}}$
(combination of $N_s^{M_S}$  in $2$) possibilities. As an example, if
$M_S = 2$ and $N_s=2$, we have $C_2^4= 6$ possibilities. Then, we store the smallest metric ($\mathcal{D}^u_{\min SR_i}$), for being
critical (a bottleneck) in terms of performance, and thus each relay
will have a minimum distance associated with its $SR$ channels. In the second step we calculate the metrics $\mathcal{D}^u_{R_iD}$
related to the $RD$ channels of each submatrix $\mathbf{H}^u_{R_i,D}$ of each relay $R_i$:
\begin{eqnarray}
    \mathcal{D}^u_{R_iD}=  \norm{\sqrt{ \frac{E}{M_S}} \mathbf{H}^u_{R_i,D}\mathbf{x}_l - \sqrt{\frac{ E}{M_S}} \mathbf{H}^u_{R_i,D}\mathbf{x}_n}^2,
    \label{eq:8}
\end{eqnarray}
where $l$ $\neq$ $n$. This metric is also calculated for each of the $C_2^{N_s^{M_S}}$  possibilities. Then, we store the minimum distance ($\mathcal{D}^u_{\min R_iD}$), and thus each submatrix $\mathbf{H}^u_{R_i,D}$ will have a minimum distance associated with its $RD$ channels. In the third step, we find the largest minimum distance $\mathcal{D}_{\min R_iD}$, and thus each relay will have its best channel submatrix $\mathbf{H}^u_{R_i,D}$ which is associated with this distance:\\
\begin{eqnarray}
 \mathcal{D}_{\min R_iD} = \max{(\mathcal{D}^u_{\min R_iD})}.
\end{eqnarray}

In the fourth step, after calculating the metrics $\mathcal{D}^u_{\min
SR_i}$  and $\mathcal{D}_{ \min R_iD}$ for each of the relays, as
described previously, we compute the expected values of $\mathcal{D}^u_{\min SR_i}$ and $\mathcal{D}_{\min R_iD}$ and adjust the $\mathcal{D}^u_{\min SR_i}$ values to balance the number of time slots selected for Max-Link-$SR$ and Max-Link-$RD$ modes:
\begin{eqnarray}
\mathcal{D}_{\min SR_i}= \frac{\mbox{E}[\mathcal{D}_{\min R_iD}]}{\mbox{E} [\mathcal{D}^u_{\min SR_i}]}\mathcal{D}^u_{\min SR_i}.
\end{eqnarray}

Then, we perform ordering and select the largest value of these distances:
\begin{eqnarray}
    \mathcal{D} _{\max \min SR-RD}= \max(\mathcal{D}_{\min SR_i}, \mathcal{D}_{\min R_iD}).
    \label{eq:9}
    \end{eqnarray}

Therefore, we select the relay that is associated with $\mathcal{D} _{\max\min
SR-RD}$, considering its buffer status. This relay will be selected for reception (if its buffer is not full) or transmission (if its buffer is not empty),
depending on this metric  is associated with the $SR$ or $RD$ channels,
respectively. Otherwise, the algorithm checks if the next maximum minimum distance and the associated relay meet the necessary requirements related to the buffer status.

\subsection{Calculation of the metric for direct transmission}

In this step we calculate the metric $\mathcal{D}_{SD}$ related to
the  $SD$ channels for the DT mode:
\begin{eqnarray}
    \mathcal{D}_{SD}=  \norm{\sqrt{\frac{2 E}{M_S}} \mathbf{H}_{S,D}\mathbf{x}_l - \sqrt{\frac{2 E}{M_S}} \mathbf{H}_{S,D}\mathbf{x}_n}^2,
    \label{eq:10}
\end{eqnarray}
where $l$ $\neq$ $n$. This metric is calculated for each
of the $C_2^{N_s^{M_S}}$  possibilities. Then, we store the minimum distance ($\mathcal{D}_{\min SD}$).
Considering imperfect CSI, the estimated channel matrix $\mathbf{\hat{H}}$ is assumed, instead of $\mathbf{H}$ in (\ref{eq:7}), (\ref{eq:8})  and (\ref{eq:10}).
After  finding $\mathcal{D}_{\max \min SR-RD}$ and $\mathcal{D}_{\min
SD}$, we compare these parameters and select the transmission mode that is equal to\\\\
$
\begin{cases}
  \mbox{Max-Link-}\textit{SR},& \mbox{if}\left(\mathcal{D} _{\max \min}= \max{(\mathcal{D}_{\min SR_i})}\right)\& \left(G> \mathcal{S}\right),\\
    \mbox{Max-Link-}\textit{RD},&\mbox{if}\left(\mathcal{D} _{\max \min}= \max{(\mathcal{D}_{\min R_i D})}\right)\& \left( G> 1\right),\\
 \mbox{DT},&\mbox{otherwise.}
            \end{cases}
$
where $\mathcal{D} _{\max \min}=\mathcal{D} _{\max \min SR-RD}$, $G=\frac{\mathcal{D}_{\max\min}}{\mathcal{D}_{\min SD}}$, and $\mathcal{S} \in \{0,1,2,\dots\}$ is a parameter that works as a switch. In \cite{f27}, assuming symmetric channels and applications without critical delay constraints, the switch $\mathcal{S}$ is equal to one. If we consider asymmetric channels and the need for a short average delay,  we select  an $\mathcal{S}$ that takes for granted that the protocol achieves a good BER and average delay performance.  If $\mathcal{S}$ is equal to zero, the protocol is selected to operate only in  the Max-Link mode and we do not have the possibility of a direct $SD$  connectivity and, consequently, we have another scheme called "MMD-Max-Link". Otherwise, when we increase $\mathcal{S}$, the number of time slots in which the protocol is selected to operate in the DT mode increases.

\section{Analysis of MMD: Impact on Relay Selection, PEP, Complexity, Sum-rate and Average Delay}

In this section, we first analyze the proposed MMD and the existing QN relay selection
 criteria. We compare
the PEP and the computational complexity of the MMD criterion versus the QN criterion.
We then derive expressions to compute the sum-rate and the average delay of the Swiched Max-Link protocol.
\subsection{Impact of the MMD and QN criteria on relay selection}

 The metrics $\mathcal{D}$ ($\mathcal{D}^u_{SR_i}$, $\mathcal{D}^u_{R_iD}$ and $\mathcal{D}_{SD}$)
are calculated in (\ref{eq:7}),  (\ref{eq:8}) and (\ref{eq:10}), for each of the $C_2^{N_s^M}$
possibilities. However, in the following, we will show that it is not necessary to calculate all these possibilities. The total number of calculations of the metric $\mathcal{D}$, needed by the MMD criterion, depends on the number $M_S$ of antennas at $S$ and $D$ and the number $M_R$ of antennas at each relay. Furthermore, it depends on the constellation (BPSK, QPSK, 16-QAM...), specifically on the number of different distances between the constellation symbols.
 For the MMD criterion to compute the metric $\mathcal{D}$,  it is necessary to consider the absolute value of the distances between the constellation symbols ($d_{c}$).  If we have BPSK and unit power symbols, $d_{c}=2$. Otherwise, if we have QPSK,  there are $W=3$ different values for $d_{c}$: $d_{c_1}=\sqrt{2}$, $d_{c_2}=\sqrt{2}j$ and $d_{c_3}=\sqrt{2}+\sqrt{2}j$.

We may consider that the $M_S \times  M_S$  channel matrix $ \mathbf{H}^u$  represents $\mathbf{H}^u_{S,R_i}$, $\mathbf{H}^u_{R_i,D}$ or $\mathbf{H}_{S,D}$. In the case of $\mathcal{D}^u_{SR_i}$ and $\mathcal{D}^u_{R_iD}$, if $\mathbf{x}_n$  and $\mathbf{x}_l$ are different from each other in just one symbol in position $j$, we have:
\begin{eqnarray}
\begin{split}
  \mathcal{D}_j&=  \norm{\sqrt{ \frac{E}{M_S}} \mathbf{H}^u\mathbf{x}_l - \sqrt{ \frac{E}{M_S}} \mathbf{H}^u\mathbf{x}_n}^2\\
&= \frac{E}{M_S}\norm{\mathbf{H}^u(\mathbf{x}_l - \mathbf{x}_n)}^2\\
&= \frac{E}{M_S}\norm{\mathbf{H}^u~ [0 \ldots \pm d_{c_w} \ldots0 ]^T}^2\\
 &=   \frac{\abs{d_{c_w}}^2 E}{M_S} \sum_{i=1}^{M_S} \abs{H^u_{i,j}}^2 \\
&~~~~~~~~~~~~~~~~~~~~~~~~~~~~~~~~~~ w=1,...,W.
  \label{eq:11}
\end{split}
\end{eqnarray}

If  $\mathbf{x}_n$  and $\mathbf{x}_l$ are different from each other in two symbols in positions $j$ and $k$, we have:
\begin{eqnarray}
\begin{split}
  \mathcal{D}_{j,k}&= \frac{E}{M_S}\norm{\mathbf{H}^u~ [0 \ldots \pm d_{c_w} \ldots \pm  d_{c_h} \ldots0 ]^T}^2\\
 &=  \frac{ E}{M_S} \sum_{i=1}^{M_S} \abs{ \pm d_{c_w} H^u_{i,j} \pm d_{c_h} H^u_{i,k}}^2 \\
&~~~~~~~~~~~~~~~~~~~~~~~~~~~~~~~~~~ w,h=1,...,W,\\
  \label{eq:12}
\end{split}
\end{eqnarray}
where the indices $w$ and $h$ may be different from each other.

If  $\mathbf{x}_n$  and $\mathbf{x}_l$ are different from each other in $M_S$ symbols, we have:
\begin{eqnarray}
\begin{split}
  \mathcal{D}_{1,...,M_S}&=  \frac{E}{M_S}\norm{\mathbf{H}^u~ [ \pm d_{c_w} \ldots \pm  d_{c_v}]^T}^2\\
&=  \frac{ E}{M_S} \sum_{i=1}^{M_S} \abs{ \pm d_{c_w} H^u_{i,1} \ldots \pm d_{c_v} H^u_{i,M_S}}^2\\
&~~~~~~~~~~~~~~~~~~~~~~~~~~~~~~~~~~ w,v=1,...,W,\\
  \label{eq:13}
\end{split}
\end{eqnarray}
where the indices $w$ and $v$ may be different from each other.

We can simplify the equations, making $\mathcal{D}=  E/M_S \times \mathcal{D'}$, where $\mathcal{D'}$=$\norm{\mathbf{H}^u(\mathbf{x}_n-\mathbf{x}_l)}^2$, for $\mathcal{D}^u_{SR_i}$ and $\mathcal{D}^u_{R_iD}$, or $\mathcal{D'}$=~2$\norm{\mathbf{H}^u(\mathbf{x}_n-\mathbf{x}_l)}^2$, for $\mathcal{D}_{SD}$. We know that the PEP considers the error event when $\mathbf{x}_n$  is transmitted and the detector computes an incorrect $\mathbf{x}_l$  (where $l\neq n$), based on the received symbol \cite{f17,f26}. If we consider $M_R=M_S$, then $U=1$ and, consequently, $\mathbf{H}=\mathbf{H}^u$  and the PEP is given by
\begin{eqnarray}
\begin{split}
\mathbf{P}(\mathbf{x}_n \rightarrow \mathbf{x}_l | \mathbf{H})&= Q\left(\sqrt{\frac{E_s}{2 N_0M_S} \mathcal{D'}}\right),
  \label{eq:19}
\end{split}
\end{eqnarray}
where $N_0$ is the  power spectrum density of the AWGN. The  MMD criterion, by maximizing the value of the minimum distance $\mathcal{D}_{\min}$, also maximizes the minimum value of the PEP argument $\mathcal{D'}_{\min}$ (PEP worst case). The PEP argument $\mathcal{D'}$  is related to the sum of the powers of the coefficients of each column (or the combination of two or more columns by addition or subtraction) of the matrix $ \mathbf{H}$. Moreover, when $U>1$, $\mathbf{H}$ is formed by multiple square submatrices $\mathbf{H}^u$, and the maximization of the minimum distances related to $\mathbf{H}^u$ also implies the maximization of the minimum value of the PEP argument.

As an example,  if we have BPSK and unit power symbols ($d_{c}=2$) and $M_S =M_R= 2$ ($U=1$), for each matrix $ \mathbf{H}^u$  ($\mathbf{H}^u_{S,R_i}$ or $\mathbf{H}^u_{R_i,D}$),  we have to calculate 4 different values for $\mathcal{D'}$:
\begin{eqnarray}
\begin{split}
\mathcal{D'}_1 &= 4~ \sum_{i=1}^{2} \abs{H^u_{i,1}}^2,~ \mathcal{D'}_2 =  4~ \sum_{i=1}^{2} \abs{H^u_{i,2}}^2,
\\
\mathcal{D'}_{1,2(+)} &=   4~ \sum_{i=1}^{2} \abs{H^u_{i,1}+H^u_{i,2}}^2,
\\~\mathcal{D'}_{1,2(-)}&=  4~ \sum_{i=1}^{2} \abs{H^u_{i,1}-H^u_{i,2}}^2.
  \label{eq:141}
\end{split}
\end{eqnarray}

If we have the direct transmission option, by considering the matrix $\mathbf{H}_{S,D}$, we also have to calculate the same expressions described in (\ref{eq:141}),  multiplied by 2. Note that these examples were considered by adopting BPSK, but other constellations (QPSK, 16-QAM...) can be adopted.

The MMD metric $\mathcal{D}$ is based on the minimum Euclidian distances between the possible vectors of transmitted symbols.  In contrast, in the QN criterion, that is  based only on the total power of these links (as the traditional Max-Link), the metric $\mathcal{Q}$ is related to the quadratic norm (the sum of the powers of all the coefficients) of each matrix $\mathbf{H}$:
\begin{eqnarray}
\begin{split}
  \mathcal{Q}&=\norm{\mathbf{H}}^2\\
&=  \sum_{j=1}^{M_S} \sum_{i=1}^{M_R} \abs{H_{i,j}}^2.\\
\end{split}
  \label{eq:14}
\end{eqnarray}

Thus, the QN criterion selects the channel matrix $\mathbf{H}^{QN}$, as given by
\begin{eqnarray}
\mathbf{H}^{QN}&= \arg \max_{\mathbf{H}}  \norm{\mathbf{H}}^2
  \label{eq:15}
\end{eqnarray}
where $\mathbf{H} \in \{\mathbf{H}_{S,R_1},\dots, \mathbf{H}_{S,R_N},\mathbf{H}_{R_1,D},\dots, \mathbf{H}_{R_N,D}\}$ and  $H_{i,j} \in \mathbb{C} (0,\sigma^2)$ .\\

The MMD criterion, differently from the QN criterion, takes into account the minimum distances related to $\mathcal{D}_j$ in (\ref{eq:11}), $\mathcal{D}_{j,k}$ in (\ref{eq:12}) and  $\mathcal{D}_{1,\dots, M_S}$ in (\ref{eq:13}), to select $\mathbf{H}^{MMD}$:
\begin{eqnarray}
\begin{split}
\mathbf{H}^{MMD}= \arg \max_{\mathbf{H}}  \min{(\mathcal{D}_j,\mathcal{D}_{j,k},\dots,\mathcal{D}_{1,\dots,M_S})}\\
~~ j, k=1,...,M_S, ~  j\neq k,
  \label{eq:16}
\end{split}
\end{eqnarray}
where $\mathbf{H} \in \{\mathbf{H}_{S,R_1},\dots, \mathbf{H}_{S,R_N},\mathbf{H}_{R_1,D},\dots, \mathbf{H}_{R_N,D},\mathbf{H}_{S,D}\}$ and  $ H_{i,j} \in \mathbb{C} (0,\sigma^2)$.\\

The advantage of the MMD algorithm as compared to QN is that MMD, by maximizing $\mathcal{D}_{\min}$, also maximizes the minimum value of the  PEP argument $\mathcal{D'}_{\min}$, whereas QN does not take it into account. So, the minimum value of the PEP argument $\mathcal{D'}_{\min}^{QN}$ associated with $\mathbf{H}^{QN}$, selected by the QN criterion,  may be not as high as the minimum value of the PEP argument $\mathcal{D'}_{\min}^{MMD}$ associated with $\mathbf{H}^{MMD}$, selected by the MMD criterion. \\

Example 1:  consider BPSK, unit power symbols and a network formed by $S$, $D$, one relay $R$ (without direct transmission), and two antennas in each node ($M_S=M_R=2$), where $\mathbf{H}_{S,R}$  and $\mathbf{H}_{R,D}$ are given by:

$$\left[
\begin{array}{c c c}
 b &2 b\\
 g & 2g\\
\end{array}\right]~ \mbox{and}~ \left[
\begin{array}{c c c}
 2b+\epsilon &2 b \\
2 g +\epsilon &  2g\\
\end{array}\right], ~ \mbox{respectively},~\mbox{where}~\epsilon \rightarrow 0.$$

By applying the QN criterion and calculating the quadratic norm of $\mathbf{H}_{S,R}$, we have: $ \mathcal{Q}=5\abs{b}^2+5\abs{g}^2$. And the quadratic norm of $\mathbf{H}_{R,D}$ is equal to: $ \mathcal{Q}=\abs{2b+\epsilon}^2+\abs{2g+\epsilon}^2+4\abs{b}^2+4\abs{g}^2. ~\mathcal{Q} \rightarrow 8\abs{b}^2+8\abs{g}^2$. Thus, by considering (\ref{eq:15}), we have: $\mathbf{H}^{QN}= \mathbf{H}_{R,D}$. In contrast, by applying the MMD criterion and calculating the minimum distance of $\mathbf{H}_{S,R}$,  we have: $\mathcal{D}_{\min}=\frac{4E}{M_S} (\abs{2b-b}^2+\abs{2g-g}^2)= \frac{4E}{M_S} (\abs{b}^2+\abs{g}^2)$. And the minimum distance of $\mathbf{H}_{R,D}$  is equal to: $\mathcal{D}_{\min}=\frac{4E}{M_S} (\abs{2b-2b-\epsilon}^2+\abs{2g-2g-\epsilon}^2)= \frac{8E}{M_S} \abs{\epsilon}^2. ~ \mathcal{D}_{\min} \rightarrow 0$.  Thus, by considering (\ref{eq:16}), we have: $\mathbf{H}^{MMD}= \mathbf{H}_{S,R}$. Moreover, by calculating the minimum values of the PEP argument, we have:  $\mathcal{D'}_{\min}^{MMD}= 4 (\abs{b}^2+\abs{g}^2)$ and $\mathcal{D'}_{\min}^{QN}=8 \abs{\epsilon}^2. ~ \mathcal{D'}_{\min}^{QN} \rightarrow0$.\\

Example 2: consider BPSK, unit power symbols and a network formed by $S$, $D$, one relay $R$ (without direct transmission), and $M_S=M_R=2$, where $\mathbf{H}_{S,R}$  and $\mathbf{H}_{R,D}$ are given by:  

$$\left[
\begin{array}{c c c}
 \epsilon_1 &5b\\
 \epsilon_2 & 4g\\
\end{array}\right]~ \mbox{and}~ \left[
\begin{array}{c c c}
 b & 3 b \\
 g & 3g\\
\end{array}\right], ~ \mbox{respectively},~\mbox{where}~\epsilon_1 \rightarrow 0~ \mbox{and}$$
$\epsilon_2 \rightarrow 0$.

By applying the QN criterion and calculating the quadratic norm of $\mathbf{H}_{S,R}$, we have: $ \mathcal{Q}=25\abs{b}^2+16\abs{g}^2+\abs{\epsilon_1}^2+\abs{\epsilon_2}^2. ~\mathcal{Q}\rightarrow 25\abs{b}^2+16\abs{g}^2$. And the quadratic norm of $\mathbf{H}_{R,D}$ is equal to: $ \mathcal{Q}=10\abs{b}^2+10\abs{g}^2$. Thus, by considering (\ref{eq:15}), we have: $\mathbf{H}^{QN}= \mathbf{H}_{S,R}$. In contrast, by applying the MMD criterion and calculating the minimum distance of $\mathbf{H}_{S,R}$,  we have: $\mathcal{D}_{\min}=\frac{4E}{M_S} (\abs{\epsilon_1}^2+\abs{\epsilon_2}^2). ~\mathcal{D}_{\min}\rightarrow 0$. And the minimum distance of $\mathbf{H}_{R,D}$  is equal to: $\mathcal{D}_{\min}=\frac{4E}{M_S} (\abs{b}^2+\abs{g}^2)$.  Thus, by considering (\ref{eq:16}), we have: $\mathbf{H}^{MMD}= \mathbf{H}_{R,D}$. Moreover, by calculating the minimum values of the PEP argument, we have:  $\mathcal{D'}_{\min}^{MMD}= 4 (\abs{b}^2+\abs{g}^2)$ and $\mathcal{D'}^{QN}_{\min}=4 (\abs{\epsilon_1}^2+\abs{\epsilon_2}^2). ~\mathcal{D'}^{QN}_{\min}\rightarrow 0$.\\

We have seen in these examples that: $\mathbf{H}^{MMD} \neq \mathbf{H}^{QN}$ and $\mathcal{D'}_{ \min}^{MMD} > \mathcal{D'}_{ \min}^{QN}$.  In the appendix, we develop a proof that shows that:
\begin{eqnarray}
\mathcal{D'}_{\min}^{MMD}\geq \mathcal{D'}_{ \min}^{QN}.
  \label{eq:18}
\end{eqnarray}

Note that these examples were considered by using BPSK, but other constellations (QPSK, 16-QAM...) can be adopted.

\subsection{Pairwise Error Probability}

As we have seen in (\ref{eq:19}), the PEP considers the error event when $\mathbf{x}_n$  is transmitted and the detector computes an incorrect $\mathbf{x}_l$  (where $l\neq n$), based on the received symbol. If we consider $M_R=M_S$, then $U=1$ and, consequently, $\mathbf{H}=\mathbf{H}^u$  and the PEP will have its maximum value for the minimum value of $\mathcal{D'}$ (worst case of the PEP).
So, for the worst case of the PEP ($\mathcal{D'}_{\min}$), in direct $SD$ transmissions, in each time slot, we have
\begin{eqnarray}
\mathbf{P}(\mathbf{x}_n \rightarrow \mathbf{x}_l | \mathbf{H})= Q\left(\sqrt{\frac{E}{2 N_0M_S} \mathcal{D'}_{\min}}\right).
  \label{eq:21}
\end{eqnarray}

However, for cooperative $SR-RD$ transmissions, an approximated expression for computing the worst case of the PEP in each time slot (regardless of whether it is an $SR$ or $RD$ link) is given by
\begin{eqnarray}
\mathbf{P}(\mathbf{x}_n \rightarrow \mathbf{x}_l | \mathbf{H})\approx 1- \left(1-Q\left(\sqrt{\frac{E}{2 N_0M_S} \mathcal{D'}_{\min}}\right)\right)^2.
  \label{eq:102}
\end{eqnarray}

The metric $\mathcal{D'}_{min}$ is maximized by the MMD criterion and the same does not happen to the QN criterion. The PEP is given by a $Q$ function and its argument is given by the root square of a constant $\left(\frac{E}{2 N_0M_S}\right)$ multiplied by $\mathcal{D'}_{\min}$. We know that by the characteristic of the $Q$ function when its argument grows its value decreases.
Therefore, if we consider (\ref{eq:18}), (\ref{eq:21}) and (\ref{eq:102}), we have
\begin{eqnarray}
\mathbf{P}^{MMD}(\mathbf{x}_n \rightarrow \mathbf{x}_l | \mathbf{H}^{MMD}) \leq \mathbf{P}^{QN}(\mathbf{x}_n \rightarrow \mathbf{x}_l | \mathbf{H}^{QN}),
  \label{eq:22}
\end{eqnarray}
where $\mathbf{P}^{MMD}(\mathbf{x}_n \rightarrow \mathbf{x}_l | \mathbf{H}^{MMD})$ is the PEP for the worst case in the MMD criterion and $\mathbf{P}^{QN}(\mathbf{x}_n \rightarrow \mathbf{x}_l | \mathbf{H}^{QN})$ is the PEP for the worst case in the QN criterion. Note that when $U>1$, $\mathbf{H}$ is formed by multiple square submatrices $\mathbf{H}^u$, and the maximization of the minimum distances related to $\mathbf{H}^u$ done by the MMD criterion also implies the maximization of the minimum value of the PEP argument.
\begin{figure}[!h]
\centering
\includegraphics[scale=0.52]{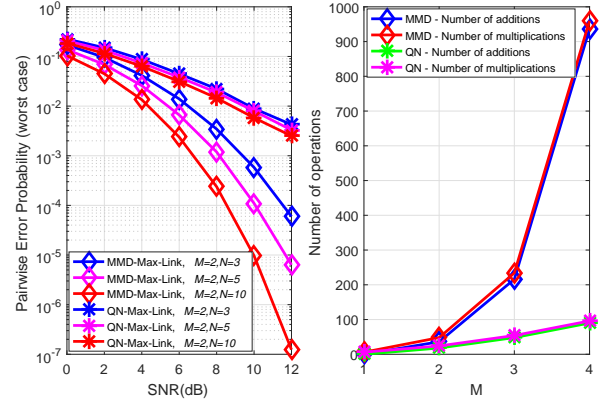}
\caption{MMD-Max-Link and QN-Max-Link a) PEP performance and b) Computational Complexity.}
\label{fig:pepMMDMaxLink}
\end{figure}

Fig. \ref{fig:pepMMDMaxLink} a) shows the theoretical PEP worst case performance (computed by the algorithm based on the selected channel matrix $\mathbf{H}$, in each time slot) of the MMD-Max-Link and QN-Max-Link protocols, for $M_S=M_R =M= 2$, $N$ = 3, 5 and 10, $J = 4$, BPSK and perfect CSI. Note that for multiple antennas the PEP worst case performance of the MMD-Max-Link scheme is much better than that of QN-Max-Link for the total range of SNR values tested. When we increase $N$, the MMD-Max-Link has its performance improved and the gap between the curves is increased. The same does not happen to QN-Max-Link,  as the QN criterion does not take the metric $\mathcal{D'}_{\min}$  into account. Note that this example was considered by adopting BPSK, but other constellations (QPSK, 16-QAM...) can be considered.

\subsection{Computational Complexity}
 We may generalize the total number  $\mathcal{X}$  of calculations of the  metric $\mathcal{D}$,  needed by the MMD criterion, for each matrix $\mathbf{H}^u_{S,R}$, $\mathbf{H}^u_{R,D}$ or $\mathbf{H}_{S,D}$:
\\
\begin{eqnarray}
\mathcal{X}= \sum_{i=1}^{M_S} 2^{i-1}  W^i C_i^{M_S},
\label{eq:23}
\end{eqnarray}
where $W$ is the total number of different distances between the constellation symbols ($d_{c}$). If we have BPSK, $W=1$, and QPSK, $W=3$. In QPSK, the calculation of some of these metrics  is redundant, so the number of calculations $\mathcal{X}$ may be less than the indicated in (\ref{eq:23}),  but it was considered in this way, by the greater ease of implementation of the algorithm.

\begin{table}[!htb]
\centering
 \caption{Computational Complexity of Criteria}
 \label{table2}
\begin{tabular}{l|ll}
\hline
Operations/Criterion& Maximum Minimum Distance & Quadratic Norm\\
\hline
additions & $2NUM_S(\mathcal{X}-1)$ & $2NU(M_S^2-1)$\\
\hline
 multiplications &  $2NUM_S\mathcal{X}$ & $2NUM_S^2$\\
\hline

\end{tabular}
\end{table}

Table \ref{table2} shows the complexity of the MMD and QN criteria for a number of $N$ relays, $M_S$ antennas at $S$ and $D$ and $M_R=UM_S$ antennas at the relays, considering only the cooperative transmission and the constellation type (BPSK, QPSK, 16-QAM...).  Fig. \ref{fig:pepMMDMaxLink} b) also shows the complexity of the MMD and QN criteria, for $N=3$ ($S$, $D$ and 3 relays), $M_S=M_R=M$ antennas at each node and BPSK. This result shows that the complexity of the MMD criterion with $M_S=2$  is not much higher than the complexity of the QN criterion. If we increase the number of antennas to  $M_S=3$ (or more)  in each node, the complexity of MMD becomes considerably higher than the complexity of QN.

\subsection{Sum-Rate}
The sum-rate of a given system is upper bounded by the system capacity. In
this context, the capacity of the cooperative system in a given time slot, using a single relay selection scheme is given by \cite{f1, f15}:
\begin{eqnarray}
  C_{DF}=\frac{1}{2} \min \{ I_{DF}^{SR},I_{DF}^{RD}\},
  \label{eq:24}
\end{eqnarray}
where the first term in (\ref{eq:24}) represents the maximum rate at which the relay can reliably decode the message from $S$,
while the second term in  (\ref{eq:24}) is the maximum rate at which $D$ can reliably decode the estimated message from $S$ transmitted by the relay \cite{f1}.

Note that in the Switched Max-Link and MMD-Max-Link schemes, differently from a single relay scheme, the selected relay for reception $R_k$ may be different from the selected relay for transmission $R_j$.  Therefore, the capacity of the MMD-Max-Link and the Switched Max-Link (operating in the Max-Link mode) is given by
\begin{eqnarray}
  C_{DF}=\frac{1}{2} \min \{ I_{DF}^{SR_k},I_{DF}^{R_jD}\},
  \label{eq:150}
\end{eqnarray}
where the first term in (\ref{eq:150}) is the maximum rate at which $R_k$ can reliably decode the message from $S$, while the second term in  (\ref{eq:150}) is the maximum rate at which $D$ can reliably decode the estimated message from $S$ transmitted by $R_j$. The capacity of direct transmission is given by
\begin{eqnarray}
  C_{DT}=I_{DT}^{SD}.
  \label{eq:25}
\end{eqnarray}

As Switched Max-Link may operate in both transmission modes (Max-Link or
 DT), the expected  sum-rate $\mathcal{R}$  in bits/Hz of this scheme, considering symmetric channels, may be expressed as:
$C_{DF}\leq \mathcal{R}\leq C_{DT}$. The relationship between mutual information and entropy can be expanded as follows for a given $\mathbf{H}_{S,R_k}$  (channel matrix from $S$ to $R_k$):
\begin{eqnarray}
\begin{split}
 I_{DF}^{SR_k}&=I_{DF}(\mathbf{x};\mathbf{y}_{S,R_k} | \mathbf{H}_{S,R_k})\\
&=\mathcal{H}(\mathbf{y}_{S,R_k})-\mathcal{H}(\mathbf{y}_{S,R_k}|\mathbf{x})\\
&=\mathcal{H}(\mathbf{y}_{S,R_k})-\mathcal{H}(\mathbf{H}_{S,R_k}\mathbf{x}+\mathbf{n}_{R_k}|\mathbf{x})\\
&=\mathcal{H}(\mathbf{y}_{S,R_k})-\mathcal{H}(\mathbf{n}_{R_k}),\\
  \label{eq:26}
\end{split}
\end{eqnarray}
where $\mathcal{H}$(·) denotes the differential entropy of a continuous random variable.  It is assumed that the transmit vector $\mathbf{x}$ and the noise vector $\mathbf{n}_{R_k}$ are independent.

Eq. (\ref{eq:26}) is maximized when $\mathbf{y}_{S,R_k}$ is Gaussian, since the normal distribution maximizes the entropy for a given variance.  For a complex Gaussian vector $\mathbf{y}_{S,R_k}$, the differential entropy is less than or equal to $\log_2 \det(\pi e\mathbf{K})$, with equality if and only if $\mathbf{y}_{S,R_k}$  is a circularly symmetric complex Gaussian vector with $ E[\mathbf{y}_{S,R_k} \mathbf{y}_{S,R_k}^H]=\mathbf{K}$ \cite{f15,f155}.  By assuming the optimal Gaussian distribution for the transmit vector $\mathbf{x}$, the covariance matrix of $\mathbf{y}_{S,R_k}$  is given by
\begin{eqnarray}
\begin{split}
E[\mathbf{y}_{S,R_k} \mathbf{y}_{S,R_k}^H]&=E[( \mathbf{H}_{S,R_k}\mathbf{x}+\mathbf{n}_{R_k})( \mathbf{H}_{S,R_k}\mathbf{x}+\mathbf{n}_{R_k})^H]\\
&= E[ \mathbf{H}_{S,R_k}\mathbf{x}(\mathbf{x})^H\mathbf{H}_{S,R_k}^H+\mathbf{n}_{R_k}(\mathbf{n}_{R_k})^H]\\
&= \mathbf{H}_{S,R_k} \mathbf{Q}_{S,R_k}\mathbf{H}_{S,R_k}^H+E[\mathbf{n}_{R_k}(\mathbf{n}_{R_k})^H]\\
&= \mathbf{H}_{S,R_k} \mathbf{Q}_{S,R_k}\mathbf{H}_{S,R_k}^H+\mathbf{K}^n\\
&=\mathbf{K}^d+\mathbf{K}^n,
  \label{eq:28}
\end{split}
\end{eqnarray}
where $d$ and $n$ denotes respectively the signal part and the noise part of (\ref{eq:28}) \cite{f155}.  The maximum mutual information is then given by

\begin{eqnarray}
\begin{split}
 I_{DF}^{SR_k}&=\mathcal{H}(\mathbf{y}_{S,R_k})-\mathcal{H}(\mathbf{n}_{R_k})\\
&=\log_2 \det(\pi e (\mathbf{K}^d+\mathbf{K}^n))-\log_2 \det (\pi e\mathbf{K}^n))\\
&=\log_2 \det(\mathbf{K}^d+\mathbf{K}^n)-\log_2 \det (\mathbf{K}^n)\\
&=\log_2 \det(\mathbf{K}^d (\mathbf{K}^n)^{-1}+\mathbf{I}_{M_R})\\
&=\log_2 \det(\mathbf{H}_{S,R_k} \mathbf{Q}_{S,R_k}\mathbf{H}_{S,R_k}^H (\mathbf{K}^n)^{-1}+\mathbf{I}_{M_R})\\
&=\log_2 \det\left(\mathbf{H}_{S,R_k} (\mathbf{Q}_{S,R_k}/N_0)\mathbf{H}_{S,R_k}^H+\mathbf{I}_{M_R}\right).\\
  \label{eq:28b}
\end{split}
\end{eqnarray}
where $\mathbf{Q}_{S,R_k}= E[\mathbf{x}(\mathbf{x})^H]= \frac{ E}{M_S}~\mathbf{I}_{M_S}$ is the covariance matrix of the transmitted symbols, $\mathbf{I}_{M_S}$ is an $M_S\times M_S$  identity matrix and $\mathbf{I}_{M_R}$ is an $M_R\times M_R$  identity matrix. Note that the vectors $\mathbf{x}$ are formed by independent and identically distributed (i.i.d.) symbols. The same reasoning can be applied to $I_{DF}^{R_jD}$ and $I_{DT}^{SD}$:
\begin{eqnarray}
I_{DF}^{R_jD}=\log_2 \det(\mathbf{H}^u_{R_j,D} (\mathbf{Q}_{R_j,D}/N_0)(\mathbf{H}_{R_j,D}^{u})^H+\mathbf{I}_{M_S}),
  \label{eq:33}
\end{eqnarray}
where $\mathbf{Q}_{R_j,D}=\frac{ E}{M_S}~\mathbf{I}_{M_S}$ and $\mathbf{H}^u_{R_j,D}$  is the selected channel submatrix from $R_j$ to $D$.
\begin{eqnarray}
I_{DT}^{SD}= \log_2 \det\left(\mathbf{H}_{S,D}(\mathbf{Q}_{S,D}/N_0)\mathbf{H}_{S,D}^H+\mathbf{I}_{M_S}\right),
  \label{eq:34}
\end{eqnarray}
where $\mathbf{Q}_{S,D}=\frac{2E}{M_S}~\mathbf{I}_{M_S}$.
For simplicity, to compute the sum-rate of the Switched Max-Link scheme, instead of considering  (\ref{eq:150}),  we considered an approximated expression for the sum-rate in each time slot, depending on the kind of transmission. Therefore, in the case of a time slot $i$ selected for $SR$ transmission, the  approximated sum-rate is given by
\begin{eqnarray}
\mathcal{R}_i^{SR_k}\approx\frac{1}{2}~ \log_2 \det\left(\mathbf{H}_{S,R_k} (\mathbf{Q}_{S,R_k}/N_0)\mathbf{H}_{S,R_k}^H+\mathbf{I}_{M_R}\right).
  \label{eq:35}
\end{eqnarray}

Furthermore, in the case of a time slot $i$ selected for $RD$ transmission, the  approximated sum-rate is given by
\begin{eqnarray}
\mathcal{R}_i^{R_jD}\approx\frac{1}{2}~  \log_2 \det(\mathbf{H}^u_{R_j,D} (\mathbf{Q}_{R_j,D}/N_0)(\mathbf{H}_{R_j,D}^{u})^H+\mathbf{I}_{M_S}).
  \label{eq:36}
\end{eqnarray}

In the case of a time slot $i$ selected for $SD$ transmission, the approximated  sum-rate is given by
\begin{eqnarray}
\mathcal{R}_i^{SD}\approx  \log_2 \det\left(\mathbf{H}_{S,D} (\mathbf{Q}_{S,D}/N_0)\mathbf{H}_{S,D}^H+\mathbf{I}_{M_S}\right).
  \label{eq:37}
\end{eqnarray}

Therefore, the average sum-rate ($\mathcal{R}$)  of the Switched Max-Link scheme
can be approximated to
\begin{eqnarray}
\mathcal{R}\approx \frac{\sum_{i=1}^{n_{SR}} \mathcal{R}_i^{SR_k}+\sum_{i=1}^{n_{RD}} \mathcal{R}_i^{R_jD}+2 \sum_{i=1}^{n_{SD}}  \mathcal{R}_i^{SD}}{n_{SR}+n_{RD}+2 n_{SD}},
  \label{eq:38}
\end{eqnarray}
where $n_{SR}$ and $n_{RD}$ represent the total number of time slots selected for transmission from $S$ to $R_k$ and from $R_j$ to $D$, respectively, in the Max-Link operation mode ($n_{SR}\cong n_{RD}$), and $n_{SD}$  is the total number of time slots selected for transmission from $S$ to $D$, in DT mode. \\

\subsection{States of buffers, outage probability and throughput}
In \cite{f9}, a framework based on Discrete Time Markov Chains
(DTMC) is proposed to analyze the traditional Max-Link algorithm, considering single-antenna systems.
This framework has been used in many subsequent works to analyze other buffer-aided relay selection protocols
whose buffer is finite \cite{f23}. In the following, we use this framework to analyze the MMD-Max-Link and the Switched Max-Link protocols for multiple-antenna systems.

The states of the DTMC  represent all
the possible states of the buffers, for both MMD-Max-Link and Switched Max-Link protocols, and also the state of direct link $SD$, for  Switched Max-Link. So, in the Switched Max-Link protocol, the transitions between the states are given by the probabilities
of successful transmissions of packets and a state of the DTMC is represented not only by the
number of sets of $M_S$ packets stored in each buffer (as in the MMD-Max-Link), but it also includes a state
which depicts the reception of $M_S$  packets directly from $S$
at $D$, denoted by $\mathcal{E}_d$ \cite{f11}. This state $\mathcal{E}_d \in \{0, 1\}$ changes
 every time a set of $M_S$ packets is received directly from $S$. If $\mathcal{E}_d$ is in state 1 and $D$ receives a set of $M_S$ packets directly from $S$ then it moves to state 0, and vice versa. Note that the state
$\mathcal{E}_d$ does not change if a set of $M_S$ packets is received by a relay, or by $D$ from a relay node.

In the Switched Max-Link protocol, the state of the DTMC can be represented by
\begin{eqnarray}
\mathcal{E}_r=(\mathcal{E}_d B_1^{r}B_2^{r}\dots B_N^{r}), ~ r\in \mathbb{N}_+, 1\leq r\leq \left(L+1\right)^N,
  \label{eq:35}
\end{eqnarray}
where $L=\frac{J}{M_S}$. The states are predefined in a random way as all the possible
$\left(L+1\right)^N$ combinations of the buffer sizes combined with the
 $\mathcal{E}_d$ state \cite{f11}.
We consider that $\mathbf{A}\in \mathbb{R}^{2\left(L+1\right)^N\times2\left(L+1\right)^N}$ denotes the state transition matrix
of the DTMC \cite{f11}, in which the entry $A_{i,j}=P(\mathcal{E}_i\rightarrow \mathcal{E}_j)=P(\mathcal{E}_{t+1}=\mathcal{E}_j|\mathcal{E}_t=\mathcal{E}_i)$
is the transition probability to move from state $\mathcal{E}_i$ at time $t$ to
state $\mathcal{E}_j$ at time ($t$+1). In order to construct the state transition
matrix $\mathbf{A}$, we have to identify the connectivity between the
different states of the buffers \cite{f9,f11}. For each time slot, the buffer
and the $\mathcal{E}_d$ status can be modified as follows: (a) the number of packets stored
in a relay buffer can be decreased by $M_S$, if a relay node
is selected for transmission in Max-Link mode (and the system is not in outage), changing the buffer status,
(b) the number of  packets stored
in a relay buffer can be increased by
$M_S$, if $S$ is selected for transmission in Max-Link mode (and the system is not in outage), changing the buffer status, (c) if $S$ is selected for transmission in DT mode (and the system is not in outage), changing the $\mathcal{E}_d$ status, (d) the buffer and the $\mathcal{E}_d$ status remain unchanged when there is an outage event (all the $SR$, $RD$ and $SD$ links in outage).

As the buffer of
each relay is finite, the DTMC can be shown to be stationary,
irreducible and aperiodic (SIA) \cite{f23,f177}.  In the following, analytical
expressions are derived for the outage probability, average throughput and
average packet delay.

An outage event occurs only when there is no change in the buffer and $\mathcal{E}_d$  status. Hence,
the outage probability of the system is given by the sum of the
product of the probabilities of being at a stage $r$ and having
an outage event \cite{f9,f11}, as given by
\begin{eqnarray}
P_{outage}=\sum_{r=1}^{Z\left(L+1\right)^N}\pi_r \bar{p}_r=\diag(\mathbf{A})\pi,
  \label{eq:35out}
\end{eqnarray}
where $Z=1$ and $Z=2$ in the MMD-Max-Link and  Switched Max-Link protocols, respectively.
By considering the MMD-Max-Link and the Swiched Max-Link (operating in Max-Link mode), if there is only
one transmission per time-slot, the average
data rate $\rho$ is 0.5 since two hops are required to reach $D$. Otherwise, in schemes with successive transmissions,
$\rho$ is approaching 1  \cite{f23}.
The proportion of the packets that make it
through is ($1-P_{outage}$). Thus, the average throughput is given
by $E[T]=\rho(1-P_{outage})$\cite{f23},
where $\rho \in (0.5, 1)$. Note that if the
links are i.i.d., then the average throughput of a relay $R_n$ \cite{f23} in the MMD-Max-Link protocol  is
given by
\begin{eqnarray}
E[T_n]=\frac{\rho(1-P_{outage})}{N}.
  \label{eq:35at}
\end{eqnarray}
And the average throughput of  $R_n$ in the Switched  Max-Link protocol  is
given by
\begin{eqnarray}
E[T_n]=\frac{\rho_{SML}(1-P_{outage})}{N},
  \label{eq:35ats}
\end{eqnarray}
where $\rho_{SML}=\frac{2\rho P^{\mathcal{S}'}_{ML}}{P^{\mathcal{S}'}_{ML}+1}$,  and $P^{\mathcal{S}'}_{ML}$ is the probability of a packet being transmitted in the Max-Link mode (passing by the relays) for a given $\mathcal{S'}$, considering $\mathcal{S'}=1$, if $\mathcal{S}\ge 1$, and $\mathcal{S'}=\mathcal{S}$, if $\mathcal{S}< 1$.

\subsection{Average Delay}

Similarly to the traditional Max-Link \cite{f9}, Switched Max-Link and MMD-Max-Link were originally considered for applications without critical delay constraints. In this work, by considering the importance of a short average delay in most modern applications, an expression for the average delay of the proposed Switched Max-Link protocol is presented.
The average delay is calculated by considering
the time a packet needs to reach the destination once it has
left the source (no delay is
measured when the packet resides at $S$ \cite{f11}). In the Switched Max-Link protocol, the direct transmission is considered to
have no delays and for packets that are processed by the relays, the
delay is the number of time slots the
packet stays in the buffer of the relay \cite{f11}.

For i.i.d. channels, the average delay is the same on all relays.
Hence, it is enough to analyze the average delay on a single
relay \cite{f23}. By Little’s law, the average packet delay at $R_n$,
denoted by $E[d_n]$ is given by
\begin{eqnarray}
E[d_n]=\frac{E[L_n]}{E[T_n]},
  \label{eq:35t}
\end{eqnarray}
where $E[L_n]$ and $E[T_n]$ are the average queue length and
average throughput, respectively \cite{f23}. So, the average queue length at
$R_n$, in the MMD-Max-Link and  Switched Max-Link protocols,  is given by
\begin{eqnarray}
E[L_n]=\sum_{r=1}^{(L+1)^N}\pi_r B_n^r.
  \label{eq:35q}
\end{eqnarray}
And the average throughput is given in (\ref{eq:35at}). Thus, by substituting (\ref{eq:35out}), (\ref{eq:35at}) and (\ref{eq:35q})
 into (\ref{eq:35t}), we have that the average delay in the MMD-Max-Link protocol  is given by

\begin{eqnarray}
E[d_n]^{MMD}=\frac{N\sum_{r=1}^{(L+1)^N} \pi_rB_n^r}{\rho\left(1-\sum_{r=1}^{(L+1)^N} \pi_r\bar{p}_r\right)},
  \label{eq:35qb}
\end{eqnarray}
where $\rho=0.5$, considering one transmission per time slot. The derivation for the average delay at the
high SNR regime is given in \cite{f177}. First the throughput of each
relay is found. As the selection of a relay is equiprobable,
the average throughput at any relay $R_n$  is  $\rho / N$, where $\rho$  is the
average data rate. Since we have half-duplex links, $\rho=1/2$ and
therefore $E[T_n]=\frac{1}{2N}$.  Also, it can be shown that the
average queue length at any relay is $E[L_n]=\frac{L}{2}$. Thus, by
Little’s law, $E[d_n]^{MMD}=E[d]=NL=N\frac{J}{M_S}$. So, as either
the number of relays or the buffer size increases, the average
delay of the MMD-Max-Link algorithm increases.

As the MMD-Max-Link protocol operates only in the Max-Link mode (similarly to the traditional Max-Link, but with multiple antennas),  we consider that the average delay of MMD-Max-Link is similar to the average delay of  Max-Link.  In contrast, the average delay of Switched Max-Link is lower than that of  Max-Link, because its advantage (the possibility of operating in DT mode). The average delay of the Switched Max-Link protocol  is given by
\begin{eqnarray}
\begin{split}
E[d_n]^{SML}&=\frac{N\sum_{r=1}^{(L+1)^N} \pi_rB_n^r}{\rho_{SML}\left(1-\sum_{r=1}^{2(L+1)^N} \pi_r\bar{p}_r\right)} \times P_{ML}^{\mathcal{S}}\\
&\approx\frac{E[d_n]^{MMD}(P^{\mathcal{S}'}_{ML}+1)}{2P^{\mathcal{S}'}_{ML}}\times P_{ML}^{\mathcal{S}},
  \label{eq:35qa}
\end{split}
\end{eqnarray}
where $P_{ML}^{\mathcal{S}}$ is the probability of a packet being transmitted  in the Max-Link mode, for a given $\mathcal{S}$. When the switch $\mathcal{S}$ tends to zero, $P_{ML}^{\mathcal{S}}$ tends to one (Switched Max-Link operates only in the Max-Link mode and its average delay equals the average delay of MMD-Max-Link). Otherwise, when  $\mathcal{S}$ tends to $\infty$, $P_{ML}^{\mathcal{S}}$ tends to zero (Switched Max-Link operates only in DT mode, and its average delay tends to zero).

\section{Numerical Results}

This section illustrates and discusses the
simulation results of the proposed Switched Max-Link, the
MMD-Max-Link, the Max-Link with direct transmission capability \cite{f11}, the conventional MIMO (direct transmission, without relaying) and the Max-Link with the QN criterion (QN-Max-Link).  QN-Max-Link with a single antenna refers to the traditional Max-Link \cite{f9}. The proposed Switched Max-Link scheme is considered in a network with $N$ relays and
$M_S$ antennas at $S$ and $D$ and $M_R$ antennas at the relays.  We considered different values for the buffer size $J$ and adopted $J=4$  packets as it is sufficient to ensure a good performance. We have also adopted $M_S=1$  and $2$ antennas. Since different packets may be stored at different relays for different amounts of time, the packets transmitted by $S$ may
arrive at $D$ in an order different from the order at $S$ \cite{f5}. To restore the original order at $D$, it was necessary to insert in the preamble of each packet the order information (its position in
the binary format, ranging from 1 to the total number of packets).
We assume that the transmitted signals belong to BPSK or QPSK
constellations. The 16-QAM constellation was not included in this work because of its higher complexity.
We also assume $N_0 =1$ and $E_S = E_{R_j} = E$ (total energy transmitted).
Scenarios with asymmetric channels were also tested in order to depict the performance of the proposed Switched Max-Link and MMD-Max-Link algorithms. The transmit signal-to-noise ratio SNR ($E/N_0$) ranges
from 0 to 12 dB and the performances of the transmission schemes
were tested for 10000$M_S$  packets, each containing 100 symbols.

\subsection{Analysis accuracy validation: PEP and BER performance}

In the following we present the theoretical PEP worst case and the simulated BER performance to validate the accuracy
of our analysis related to the MMD relay selection criterion, adopted in the Switched Max-Link and the MMD-Max-Link protocols. Then, the BER, average throughput and average delay performances of the Switched Max-Link and Max-Link with direct transmission capability \cite{f11} protocols are compared.  We also present the BER performance considering BPSK, QPSK and outdated CSI of the Switched Max-Link, MMD-Max-Link and conventional MIMO protocols, considering unit power links ($\sigma_{ S,R}^2=\sigma_{ R,D}^2=\sigma_{ S,D}^2=1$).

\begin{figure}[!h]
\centering
\includegraphics[scale=0.5]{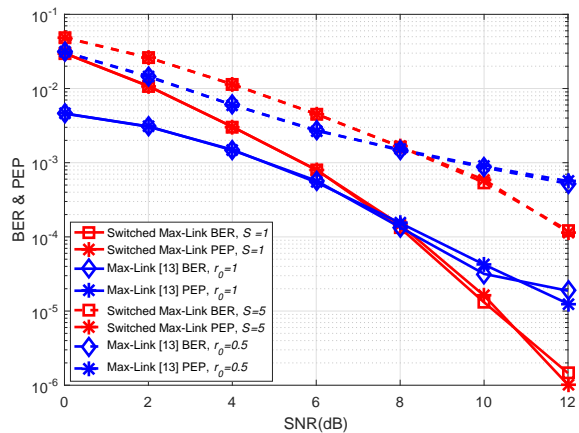}
\caption{Switched Max-Link and Max-Link \cite{f11} PEP and BER performances.}
\label{fig:PEPexMMDMaxLink}
\end{figure}

Fig. \ref{fig:PEPexMMDMaxLink} shows the theoretical PEP performance that yields from our theoretical framework that has been presented in Section IV and the BER performance of the Switched Max-Link and Max-Link \cite{f11} protocols, for BPSK, $M_S= M_R = 1$, $N$ = 3 and $J = 4$.  In Switched Max-Link, we have $\mathcal{S}=1$ (solid curve)  and 5 (dashed curve), and in Max-Link, we have $r_0=1$ (solid curve) and 0.5 BPCU (bits per channel use) (dashed curve).  By comparing the solid curves, the result shows that for low SNR values (less than 8dB), the Max-Link  protocol  has a better BER performance than that of Switched Max-Link. This is because if an outage event occurs in Max-Link, the packet is not transmitted (improving the BER, but reducing the average throughput). In contrast, Switched Max-Link has a better BER performance than that of  Max-Link for SNR values greater than 8dB, resulting also in a higher diversity gain. And the results are the same when we compare the dashed curves. These results show that the theoretical  PEP performance matches the BER performance and validate the  accuracy
of our analysis. Note that in this case we have just a pair of  possible transmitted symbols, so the BER performance is comparable to the PEP performance.

\begin{figure}[!h]
\centering
\includegraphics[scale=0.5]{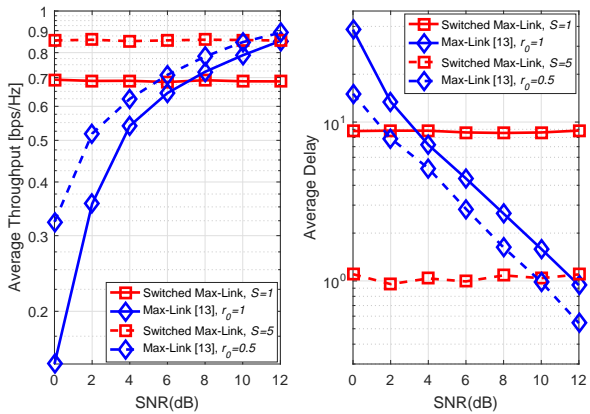}
\caption{Switched Max-Link and Max-Link \cite{f11} a) average throughput and b) average delay.}
\label{fig:ThADexMMDMaxLink}
\end{figure}

Fig. \ref{fig:ThADexMMDMaxLink}  shows the average throughput and average delay performances of the Switched Max-Link and Max-Link \cite{f11} protocols, for the same configuration described in Fig. \ref{fig:PEPexMMDMaxLink}. The Switched Max-Link protocol has a high average throughput even for low SNR values. This does not happen to Max-Link, as in this protocol, if an outage event occurs, the packet is not transmitted (reducing the average throughput). Moreover, Switched Max-Link has a low average delay (when $\mathcal{S}=5$)  even for low SNR values as opposed to Max-Link.

\begin{figure}[!h]
\centering
\includegraphics[scale=0.53]{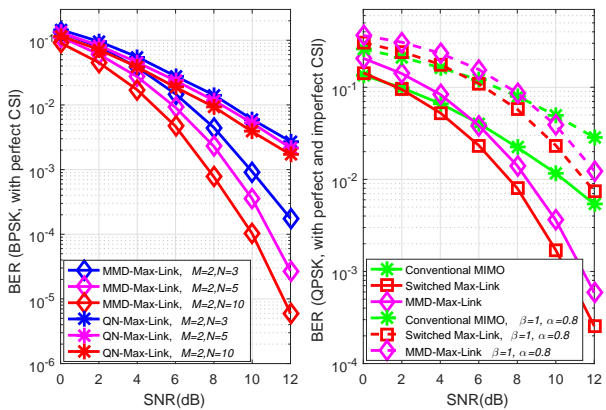}
\caption{a) BER performance for BPSK and b) BER performance for QPSK, with  perfect and imperfect channel knowledge.}
\label{fig:pepberMMDMaxLink}
\end{figure}

Fig. \ref{fig:pepberMMDMaxLink}  a) shows the BER performance of the MMD-Max-Link and QN-Max-Link protocols, for $M_S=M_R = M=2$, $N$ = 3, 5 and 10, $J = 4$, BPSK and perfect CSI. Note that for multiple antennas the BER performance of the MMD-Max-Link scheme is much better than that of QN-Max-Link for the total range of SNR values tested. When we increase $N$, the MMD-Max-Link has its performance improved. The same does not happen to QN-Max-Link,  as the QN criterion does not take the metric $\mathcal{D'}_{\min}$  into account. This result validates the accuracy of our analysis in Section IV, illustrating that a better theoretical  PEP worst case performance achieved by the MMD relay selection criterion implies also a better BER performance for the MMD-Max-Link protocol.
Fig. \ref{fig:pepberMMDMaxLink} b) shows the Switched Max-Link, the MMD-Max-Link
and the conventional MIMO BER  performance
comparison for $M_S=M_R= 2$,  $N= 10$,  $J= 4$, $\mathcal{S}=1$, QPSK, perfect and  imperfect CSI ($\beta=1$ and $\alpha=0.8$). The QN-Max-Link was not considered as its performance is worse than the performance of the proposed protocol. Both for perfect and imperfect CSI, the
performance of Switched
Max-Link is considerably better than that of the
conventional MIMO for a wide range of SNR values. Switched Max-Link also outperforms MMD-Max-Link, and has resiliency in low transmit SNR conditions. Moreover, we note that outdated CSI results in diversity loss.

\subsection{Performance under asymmetric channels}

In the following we consider the BER, sum-rate and average delay performances of the proposed and existing schemes under asymmetric channels.

\begin{figure}[!h]
\centering
\includegraphics[scale=0.5]{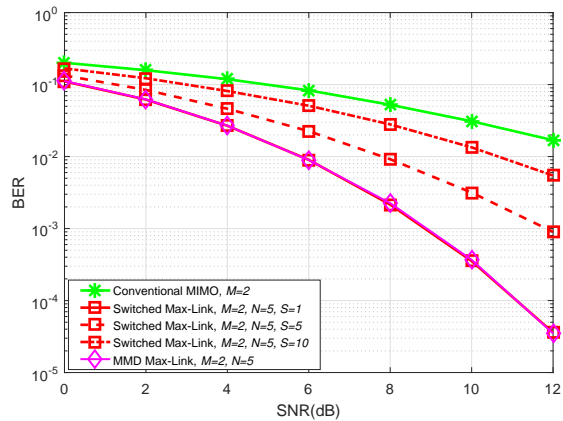}
\caption{BER performance, with low power $SD$ links.}
\label{fig:berlowsd}
\end{figure}
Fig. \ref{fig:berlowsd} shows the BER performance of the Switched Max-Link, MMD-Max-Link and the conventional MIMO protocols, for $M_S=M_R= M=2$,  $N$= 5, $J=4$, $\mathcal{S}=1 $, 5  and 10, BPSK, perfect CSI and  low power $SD$ links ($\sigma_{ S,R}^2$
$=$ $\sigma_{ R,D}^2$ $=1$  and $\sigma_{S,D}^2$ $= 0.2$). The performance of the proposed Switched Max-Link scheme, for $\mathcal{S}=1$, is very close to that of the MMD-Max-Link, illustrating the importance of switching to the Max-Link mode, when we have low power $SD$ links.

 \begin{figure}[!h]
\centering
\includegraphics[scale=0.47]{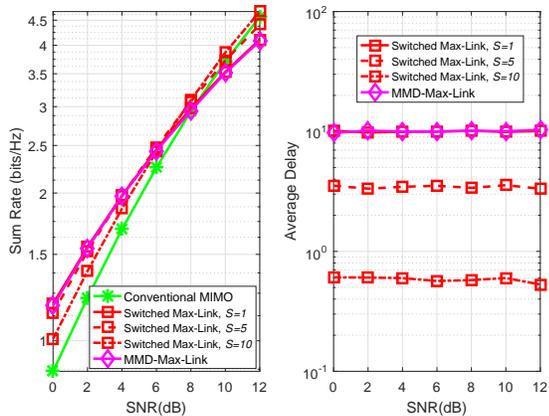}
\caption{a) Sum-rate and b) average delay performances, with low power $SD$ links.}
\label{fig:sumratelowsd}
\end{figure}

Fig. \ref{fig:sumratelowsd} shows the sum-rate (assuming Gaussian signaling) and the average delay performances of the Switched Max-Link, MMD-Max-Link and the conventional MIMO protocols, for the same configuration described in Fig. \ref{fig:berlowsd}. We notice that the simulated average delay of the MMD-Max-Link is equal to its theoretical value $\left(\frac{NJ}{M_S}=10\right)$. This result validates the  accuracy of our analysis in Section IV. When we increase $\mathcal{S}$ in the proposed Switched Max-Link, the average delay reduces and is less than 1 time slot, when  $\mathcal{S}$  is equal to 10.  This result also validates the  accuracy of our analysis. Moreover, the sum-rate of the proposed Switched Max-Link, for SNR values less than 6dB, is increased when we reduce $\mathcal{S}$, and,  for SNR values  greater than 6dB, it is increased when we increase $\mathcal{S}$.

\begin{figure}[!h]
\centering
\includegraphics[scale=0.46]{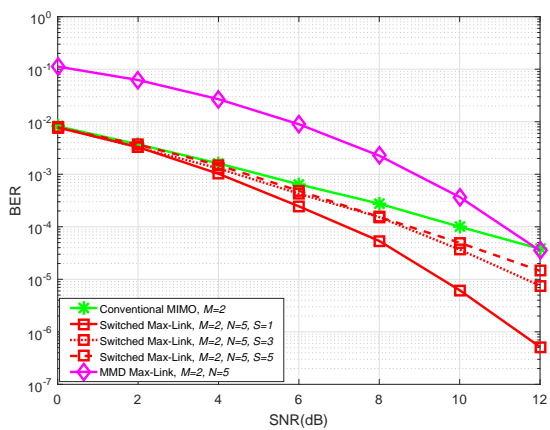}
\caption{BER performance, with high power $SD$ links.}
\label{fig:berhighsd}
\end{figure}

Fig. \ref{fig:berhighsd} shows the BER performance of the Switched Max-Link, MMD-Max-Link and the conventional MIMO protocols, for $M_S=M_R=M= 2$,  $N$= 5, $J=4$, $\mathcal{S}=1 $, 3  and 5, BPSK, perfect CSI and high power $SD$ links ($\sigma_{S,R}^2$
$=$ $\sigma_{R,D}^2$ $=1$  and $\sigma_{S,D}^2$ $=5$). The performance of the proposed Switched Max-Link scheme, for the $\mathcal{S}$ values tested, is better than that of the conventional MIMO and considerably better than that of the MMD-Max-Link scheme, illustrating the importance of switching to DT mode, when we have high power $SD$ links.

 \begin{figure}[!h]
\centering
\includegraphics[scale=0.47]{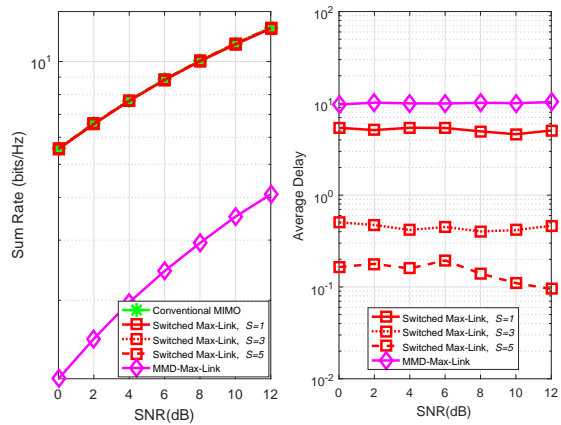}
\caption{a) Sum-rate and b) average delay performances, with high power $SD$ links.}
\label{fig:sumratehighsd}
\end{figure}

Fig. \ref{fig:sumratehighsd} shows the sum-rate and the average delay performances of the Switched Max-Link, MMD-Max-Link and the conventional MIMO protocols, for the same configuration described in Fig. \ref{fig:berhighsd}. When we increase $\mathcal{S}$ in the proposed Switched Max-Link, the average delay reduces and is less than 1 time slot, when  $\mathcal{S}$  is greater than 3. Moreover, the sum-rate performance of the proposed Switched Max-Link (for all the $\mathcal{S}$ values tested) is very close to that of conventional MIMO, for all the range of  SNR values tested, and considerably higher than that of the MMD-Max-Link scheme.

\begin{figure}[!h]
\centering
\includegraphics[scale=0.44]{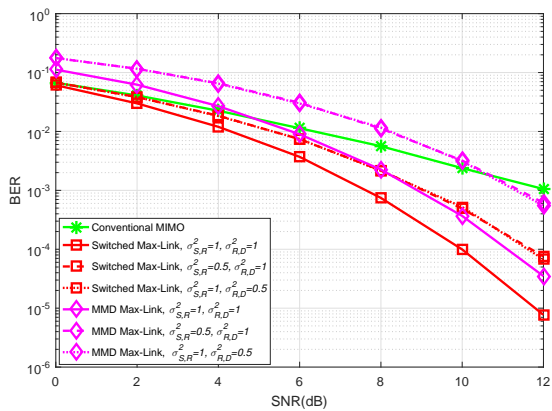}
\caption{BER performance, with low power $SR$ or $RD$ links.}
\label{fig:berlowsrrd}
\end{figure}

Fig. \ref{fig:berlowsrrd} shows the BER performance of the Switched Max-Link, MMD-Max-Link and the conventional MIMO protocols, for $M_S=M_R=M= 2$,  $N$= 5, $J=4$, $\mathcal{S}=1 $, BPSK, perfect CSI and  low power $SR$ or $RD$ links ($\sigma_{S,R}^2=0.5$ and 1, $\sigma_{R,D}^2 = 0.5$  and 1, $\sigma_{S,D}^2=1$). Switched Max-Link outperforms conventional MIMO and MMD-Max-Link schemes, illustrating that even with low power $SR$ or $RD$ links, Switched Max-Link has a better performance than that of  conventional MIMO.

 \begin{figure}[!h]
\centering
\includegraphics[scale=0.45]{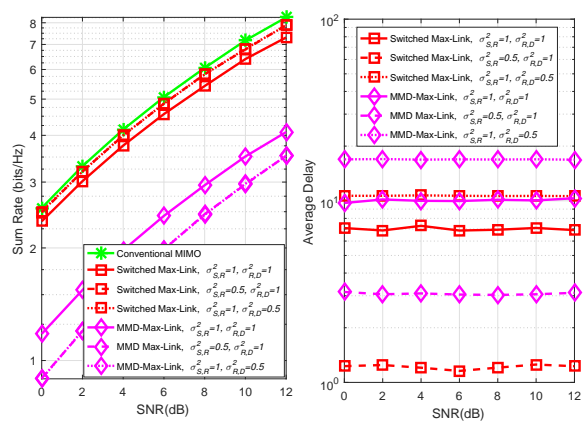}
\caption{a) Sum-rate and b) average delay performances, with low power $SR$ or $RD$ links.}
\label{fig:sumratelowsrrd}
\end{figure}

Fig. \ref{fig:sumratelowsrrd} shows the sum-rate and the average delay performances of the Switched Max-Link, MMD-Max-Link and the conventional MIMO protocols, for the same configuration described in Fig. \ref{fig:berlowsrrd}. When we have low power $SR$ links ($\sigma_{S,R}^2=0.5$ and $\sigma_{R,D}^2=1$), the probalility of selecting an $SR$ link is less than the probability of selecting an $RD$ link, so the average delay is less than the average delay with equal unit power channels ($\sigma_{S,R}^2=1$ and $\sigma_{R,D}^2=1$). Otherwise, when we have low power $RD$ links ($\sigma_{S,R}^2=1$ and $\sigma_{ R,D}^2=0.5$), the probalility of selecting an $RD$ link is less than the probability of selecting an $SR$ link, so the average delay is greater than the average delay with equal unit power channels. Moreover, the sum-rate performance of the proposed Switched Max-Link  is very close to that of conventional MIMO, even for low power $SR$ or $RD$ links, and considerably higher than that of the MMD-Max-Link scheme. The slightly
worse sum-rate performance of  Switched Max-Link compared to conventional MIMO is justified, as the proposed scheme is able to transmit with higher order modulation due to the improved BER performance.

\subsection{Performance for Massive MIMO}

In the following we consider the performance of the proposed scheme for massive MIMO (with a small number of antennas at $S$ and $D$ and a large number of antennas at the relays).
 \begin{figure}[!h]
\centering
\includegraphics[scale=0.52]{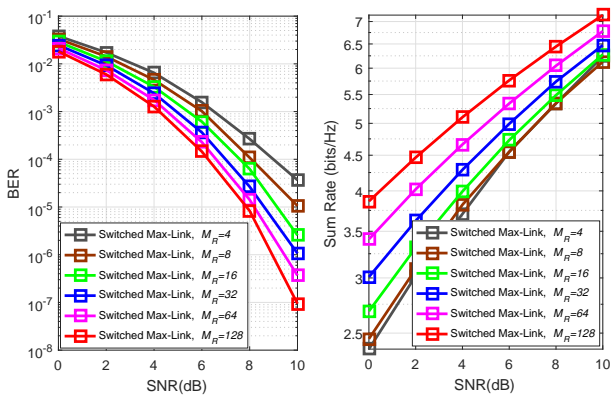}
\caption{a) BER and b) sum-rate performances, for massive MIMO.}
\label{fig:bermassive}
\end{figure}

Fig. \ref{fig:bermassive} shows the BER and sum-rate performances of the Switched Max-Link protocol, for $M_S= 2$, $M_R=4$,  8, 16, 32, 64 and 128,  $N$= 5, $J=4$, $\mathcal{S}=1 $, BPSK, perfect CSI and  unit power links ($\sigma_{ S,R}^2=\sigma_{ R,D}^2=\sigma_{ S,D}^2=1$).  Both the BER and sum-rate performances are considerably improved when we increase $M_R$, illustrating that the proposed protocol can be used for massive MIMO (with a small number of antennas at $S$ and $D$ and a  large number of antennas at the relays). This result validates  the  accuracy of our analysis, as when $U>1$, the maximization of the minimum distances related to $\mathbf{H}^u$ also implies the maximization of the minimum value of the PEP argument. Note that the achieved BER values were considerably reduced, thus the transmit signal-to-noise ratio SNR ($E/N_0$) ranges
from 0 to 10 dB.

\section{Conclusions}

In this paper, we have presented the benefits of using a novel relay selection protocol based on switching and the selection of the best link, denoted as Switched Max-Link. We then consider the MMD relay selection criterion for MIMO systems, along with algorithms that are incorporated into the proposed Switched Max-Link protocol.
Switched Max-Link was evaluated experimentally and outperformed the conventional direct transmission
and the existing QN Max-Link scheme. Despite the higher complexity of the proposed Switched Max-Link with the  MMD relay selection criterion, it is an attractive solution for relaying systems with source and destination nodes equipped with a small number of antennas and relay nodes equipped with a small or large number of antennas due to its high performance and reduced delay.

\appendices

\section{Proof of  $\mathcal{D'}_{\min}^{MMD} \geq \mathcal{D'}_{\min}^{QN}$}

The selected matrix by the MMD criterion, that maximizes the minimum distances $\mathcal{D}$, is given by
\begin{eqnarray}
\begin{split}
\mathbf{H}^{MMD}&= \arg \max_{\mathbf{H}}  \min{(\mathcal{D}_j,\mathcal{D}_{j,k},\dots,\mathcal{D}_{1,\dots,M_S})}\\
& ~~~~~~~~~~~~~~~~~~~~~~~~ j, k=1,...,M_S, ~  j\neq k,\\
\end{split}
\end{eqnarray}
where $\mathbf{H} \in \{\mathbf{H}_{S,R_1},\dots, \mathbf{H}_{S,R_N},\mathbf{H}_{R_1,D},\dots, \mathbf{H}_{R_N,D},\mathbf{H}_{S,D}\}$ and  $ H_{i,j} \in \mathbb{C} (0,\sigma^2)$ .\\

 As $\mathcal{D}=\frac{E}{M_S} \mathcal{D'}$,  where $\mathcal{D'}$=$\norm{\mathbf{H}^u(\mathbf{x}_n-\mathbf{x}_l)}^2$, for $\mathcal{D}^u_{SR_i}$ and $\mathcal{D}^u_{R_iD}$, or $\mathcal{D'}$=~2$\norm{\mathbf{H}^u(\mathbf{x}_n-\mathbf{x}_l)}^2$ , for $\mathcal{D}_{SD}$,  we have\\
\begin{eqnarray}
\begin{split}
\mathbf{H}^{MMD}= \arg &\max_{\mathbf{H}}  \min{(\mathcal{D'}_j,\mathcal{D'}_{j,k},\dots,\mathcal{D'}_{1,\dots,M_S})}\\
& ~~~~~~~~~~~~~~~~~~~ j, k=1,...,M_S, ~  j\neq k,\\
\rm{where~the~PEP}&\rm{~arguments~\mathcal{D'}~are~given~by}\\
~ \mathcal{D'}_j&=   \abs{d_{c_w}}^2  \sum_{i=1}^{M_S} \abs{H^u_{i,j}}^2\\
& ~~~~~~~~~~~~~~~~~~ w=1,...,W,\\
\mathcal{D'}_{j,k}&= \sum_{i=1}^{M_S} \abs{ \pm d_{c_w} H^u_{i,j} \pm d_{c_h} H^u_{i,k}}^2\\
& ~~~~~~~~~~~~~~~~~~ w,h=1,...,W,\\
 \mathcal{D'}_{1,\dots,M_S}&= \sum_{i=1}^{M_S} \abs{ \pm d_{c_w} H^u_{i,1}\ldots \pm d_{c_v} H^u_{i,M_S}}^2\\
& ~~~~~~~~~~~~~~~~~~ w,v=1,...,W.\\
\end{split}
\end{eqnarray}
\\

So, the maximized minimum value of the PEP argument associated to $\mathbf{H}^{MMD}$ is given by\\
\begin{eqnarray}
\begin{split}
\mathcal{D'}_{\min}^{MMD}&=\min{(\mathcal{D'}^{MMD}_j,\mathcal{D'}^{MMD}_{j,k},\dots,\mathcal{D'}^{MMD}_{1,\dots,M_S})}\\
& ~~~~~~~~~~~~~~~~~~~~~~~~~~ j, k=1,...,M_S,  ~ j\neq k.\\
\end{split}
\end{eqnarray}

However, the selected matrix by the QN criterion is given by\\
\begin{eqnarray}
\begin{split}
\mathbf{H}^{QN}&= \arg \max_{\mathbf{H}}  \sum_{j=1}^{M_S} \sum_{i=1}^{M_R} \abs{H_{i,j}}^2\\
& =\arg \max_{\mathbf{H}}\left(\sum_{i=1}^{M_R} \abs{H_{i,1}}^2+\dots+\sum_{i=1}^{M_R} \abs{H_{i,M_S}}^2\right),
\end{split}
\end{eqnarray}
where $\mathbf{H} \in \{\mathbf{H}_{S,R_1},\dots, \mathbf{H}_{S,R_N},\mathbf{H}_{R_1,D},\dots, \mathbf{H}_{R_N,D}\}$ and  $H_{i,j} \in \mathbb{C} (0,\sigma^2)$ .\\

The minimum value of the PEP argument associated to
$\mathbf{H}^{QN}$ is given by\\
\begin{eqnarray}
\begin{split}
\mathcal{D'}_{\min}^{QN}&=\min{(\mathcal{D'}^{QN}_j,\mathcal{D'}^{QN}_{j,k},\dots,\mathcal{D'}^{QN}_{1,...,M_S})}\\
& ~~~~~~~~~~~~~~~~~~~~~~~~~~ j, k=1,...,M_S, ~  j\neq k.\\
\end{split}
\end{eqnarray}

If the sum of the powers of the coefficients of
one of the columns (or the combination of 2 or more
columns by addition or subtraction) of a selected submatrix and/or matrix $\mathbf{H}^{QN}$ is very small or tends to zero, we have
\begin{eqnarray}
\begin{split}
\mathcal{D'}^{QN}_j &\rightarrow 0, \mathcal{D'}^{QN}_{j,k}\rightarrow 0,\dots, \rm{or}~\mathcal{D'}^{QN}_{1,...,M_S}\rightarrow 0,
\end{split}
\end{eqnarray}\\
and, consequently: $\mathcal{D'}_{\min}^{QN}\rightarrow 0$.\\

As MMD maximizes $\mathcal{D'}_{\min}$, the submatrix and/or matrix selected by QN will be different from the selected by MMD:\\

$\mathbf{H}^{QN}\neq\mathbf{H}^{MMD}$  and $\mathcal{D'}_{\min}^{QN} \neq \mathcal{D'}_{\min}^{MMD}.$\\

We have seen that the MMD criterion computes all the values of $\mathcal{D'}$ and stores its minimum value ($\mathcal{D'}_{\min}$), for each submatrix $\mathbf{H}^u$.  Then, this  criterion selects the matrix $\mathbf{H}$  ($\mathbf{H}^{MMD}$)  that is associated to the maximum $\mathcal{D'}_{\min}$ ($ \mathcal{D'}_{\min}^{MMD}$) and the associated relay.  As the goal of this criterion is to maximize the argument of the PEP in its worst case ($\mathcal{D'}_{\min}$), another criterion such as QN can not outperform MMD but only equalize its performance, resulting in the same $\mathcal{D'}_{\min}$,  if the matrix selected by QN ($\mathbf{H}^{QN}$) is equal to $\mathbf{H}^{MMD}$. Therefore, if we have $\mathbf{H}^{QN} \neq \mathbf{H}^{MMD}$, this implies that the $ \mathcal{D'}_{\min}$ associated to $\mathbf{H}^{MMD}$  ($ \mathcal{D'}_{\min}^{MMD}$) is greater than  the $ \mathcal{D'}_{\min}$ associated to $\mathbf{H}^{QN}$  ($ \mathcal{D'}_{\min}^{QN}$). As there are cases where $\mathbf{H}^{QN} \neq \mathbf{H}^{MMD}$, we may conclude that: $\mathcal{D'}_{\min}^{MMD} \geq \mathcal{D'}_{\min}^{QN}$.

\ifCLASSOPTIONcaptionsoff
  \newpage
\fi

\end{document}